\documentclass[12pt]{iopart}
\expandafter\let\csname equation*\endcsname=\relax
\expandafter\let\csname endequation*\endcsname=\relax
\usepackage[english]{babel}
\usepackage{lipsum}
\usepackage{amsfonts, mathtools, systeme}
\usepackage{graphicx, xcolor}
\usepackage{algorithmic}
\usepackage[caption=false]{subfig}
\usepackage{slashbox}

\usepackage{cite}
\usepackage[colorlinks=true,linkcolor=blue,citecolor=red]{hyperref}%
\usepackage{cleveref}
\begin{document}
	
	\title[Effects of target anisotropy]{Effects of target anisotropy on harmonic measure and mean first-passage time}
	
	\author{Adrien Chaigneau and Denis S. Grebenkov}
	
	\address{Laboratoire de Physique de la Matière Condensée, CNRS, Ecole Polytechnique, Institut Polytechnique de Paris, 91120 Palaiseau, France}
	\ead{adrien.chaigneau@polytechnique.edu denis.grebenkov@polytechnique.edu}
	\vspace{10pt}
	\begin{indented}
		\item[]March 2023
	\end{indented}
	
	\begin{abstract}
		We investigate the influence of target anisotropy on two characteristics of diffusion-controlled reactions: harmonic measure density and mean first-passage time. First, we compute the volume-averaged harmonic measure density on prolate and oblate spheroidal targets inside a confining domain in three dimensions. This allows us to investigate the accessibility of the target points to Brownian motion. In particular, we study the effects of confinement and target anisotropy.  The limits of a segment and a disk are also discussed. Second, we derive an explicit expression of the mean first-passage time to such targets and analyze the effect of anisotropy. In particular, we illustrate the accuracy of the capacitance approximation for small targets.
	\end{abstract}
	
	%
	%
	%
	%
	%
	\section{Introduction}
	Diffusion-controlled reactions play a prominent role in chemistry, biology and engineering applications \cite{metzler2014first, redner2001guide, rice1985diffusion}.
	Smoluchowski \cite{smoluchowski1918versuch} was the first to formalize diffusion-controlled reactions in terms of diffusion equation that governs time evolution of a concentration of particles diffusing towards a static target with an appropriate boundary condition that specifies the reactivity of the target. At a single-molecule level, the search of the  target and the consequent reaction on it are characterized by the so-called first-passage statistics. When dealing with a small target, various characteristics of diffusion-controlled reactions can be obtained explicitly, such as the mean first-passage time or the smallest eigenvalue of the governing Laplace operator \cite{cheviakov2011optimizing, kolokolnikov2005optimizing, maz1984asymptotic}. Most former studies concerned a spherical target which is characterized by a single lenghtscale (its radius). If a small sphere is replaced by a small cube or another nearly isotropic shape of the same size, the reaction rate or trapping capacity for diffusing particles do not change much \cite{grebenkov2022mean}. Yet, an anisotropic target presents at least two geometrically relevant lenghtscales: ``lenght'' and ``width'', such that it is not obvious to say what is the target size. In this light, anisotropy deserves to be studied on its own right. Despite several former studies on the impact of the target shape \cite{chang2016green, gadzinowski2021spherical, grimes2018oxygen, miller1991trapping, richter1974diffusion, torquato1991trapping,  traytak2018diffusion}, the role of target anisotropy in diffusion-controlled reactions remains poorly  understood. 
	
	In this paper, we consider a particle that starts from a point $\boldsymbol{x}_0$ and diffuses with a diffusion coefficient $D$ inside a bounded confining domain
	$\Omega \in \mathbb{R}^3$ with a smooth boundary $\partial\Omega = \partial \Omega_0 \cup \Gamma$ composed
	of two disjoint parts: a reflecting “outer” boundary $\partial \Omega_0$ and an absorbing
	“inner” part $\Gamma$, that we call a target.
	We study two characteristics of diffusion-controlled reactions: the volume-averaged harmonic measure density and the mean first-passage time. The harmonic measure $\omega(X, \boldsymbol{x}_0)$ of a subset $X$ of the absorbing boundary $\Gamma$ is the probability that a Brownian motion started from $\boldsymbol{x}_0 \in \Omega$ hits that subset first, before hitting the remaining parts $\Gamma \backslash X$ \cite{garnett2005harmonic}. For smooth boundaries, one can introduce the harmonic measure density $\omega(\boldsymbol{x}, \boldsymbol{x}_0)$,  to write 
	
	\begin{equation}
		\omega(X, \boldsymbol{x}_0) = \int_{X} \omega(\boldsymbol{x}, \boldsymbol{x}_0) d\boldsymbol{x}.
	\end{equation}
	Given a point $\boldsymbol{x} \in \Gamma$, $\omega(\boldsymbol{x}, \boldsymbol{x}_0)d\boldsymbol{x}$ is the probability of the first arrival in the vicinity $d\boldsymbol{x}$ of $\boldsymbol{x}$. The harmonic measure and its density have been thorougly investigated in mathematical and physical literature \cite{adams2009harmonic, evertsz1992harmonic, garnett1986applications, garnett2005harmonic, grebenkov2005makes, grebenkov2005multifractal}. In particular, the harmonic measure density can be obtained as 
	\begin{equation}
		\omega(\boldsymbol{x}, \boldsymbol{x}_0) = -\partial_n G(\boldsymbol{x}, \boldsymbol{x}_0),
	\end{equation}
	where $\partial_n$ is the normal derivative oriented outwards the domain, and $G(\boldsymbol{x}, \boldsymbol{x}_0)$ is the Green's function which satisfies:
	
	\begin{equation}
		\left\{ \begin{array}{r l l}\label{prob:G}
			- \Delta G(\boldsymbol{x}, \boldsymbol{x}_{0} )= &\delta(\boldsymbol{x}-\boldsymbol{x}_{0}) &\quad (\boldsymbol{x} \in \Omega), \\
			G(\boldsymbol{x}, \boldsymbol{x}_{0})= &0 &\quad (\boldsymbol{x} \in \Gamma), \\
			\partial_{n} G(\boldsymbol{x}, \boldsymbol{x}_{0}) = &0 &\quad (\boldsymbol{x} \in \partial \Omega_{0}),
		\end{array} \right.
	\end{equation}
	where $\delta\left(\boldsymbol{x}-\boldsymbol{x}_{0}\right)$ is the Dirac distribution, and $\Delta$ the Laplace operator. 
	
	In this paper, we focus on the volume-averaged harmonic measure density
	\begin{equation}
		\omega(\boldsymbol{x}) = \frac{1}{|\Omega|}\int_{\Omega} \omega(\boldsymbol{x}, \boldsymbol{x}_0)d\boldsymbol{x}_0,
	\end{equation}
	i.e., the average over the starting point $\boldsymbol{x}_0$, as if it was uniformly distributed in the confining domain of volume $|\Omega|$. The volume-averaged harmonic measure density is then
	
	\begin{equation}\label{eq:defw}
		\omega(\boldsymbol{x}) = \frac{1}{|\Omega|} \int_{\Omega} \left(- \partial_{n} G\left(\boldsymbol{x}, \boldsymbol{x}_{0}\right)\right) d \boldsymbol{x}_{0}.
	\end{equation}
	
	In the simple case when $\Gamma$ and $\partial \Omega_0$ are two concentric spheres, the rotational symmetry of the domain  implies that the volume-averaged harmonic measure density is uniform, i.e. all targets points are equally accessible to Brownian motion.
	One may wonder if the uniformity holds approximately in the general case of a small target of arbitrary shape. As the starting point is distributed uniformly inside the domain, one may expect that the diffusing particle  has almost equal probabilities to reach different parts of the target despite its anisotropy.
	%
	Such uniformity was one of the assumptions in our previous work  \cite{chaigneau2022first}. We will discuss the validity of this assumption for anisotropic targets. 
	
	We also consider the mean first-passage time $T(\boldsymbol{x}_0)$ to the target $\Gamma$ from a starting point $\boldsymbol{x}_0$. The mean first-passage time satisfies the boundary value problem
	
	\begin{equation}\label{prob:T}
		\left\{ \begin{array}{r l}
			-D \Delta T(\boldsymbol{x}_0) = 1 & \quad (\boldsymbol{x}_0 \in \Omega), \\
			T(\boldsymbol{x}_0) = 0 &\quad  (\boldsymbol{x}_0 \in \Gamma), \\
			\partial_n T(\boldsymbol{x}_0) = 0 &\quad (\boldsymbol{x}_0 \in \partial \Omega_0).
		\end{array}  \right.
	\end{equation}
	As a consequence, it can be expressed in terms of the Green's function as  
	\begin{equation}\label{eq:defT}
		T(\boldsymbol{x}_0) = \frac{1}{D}\int_{\Omega} G(\boldsymbol{x}, \boldsymbol{x}_0)d\boldsymbol{x}.
	\end{equation}
	One may wonder how target anisotropy affects the mean first-passage time, or to what extent the mean first-passage time to an anisotropic target is different from that to a spherical target.
	The present work aims to answer these questions and thus to complete our knowledge on the effect of target anisotropy in diffusion-controlled reactions.
	
	The paper is organized as follows. Section \ref{sec:prolate} is devoted to the effect of anisotropy of elongated targets, modeled by prolate spheroids. We start by recalling the prolate spheroidal coordinates, then we study the volume-averaged harmonic measure density for elongated targets. We also investigate the effect of anisotropy of the target on the mean first-passage time for prolate spheroids. In particular,  we compare the mean first-passage time towards a spheroidal target and an ``equivalent'' spherical target.
	Section \ref{sec:oblate} follows the same structure for flattened targets, modeled by oblate spheroids.
	In section \ref{sec:discussion}, we discuss the results and conclude. Appendix contains some technical details. 

	\section{Elongated targets}\label{sec:prolate}
	In this section, we consider the domain $\Omega$ between biaxial concentric prolate spheroids in three dimensions. After recalling the prolate spheroidal coordinates, we obtain the volume-avegared harmonic measure density and investigate the effect of anisotropy on the mean first-passage time in such domains.\\
	\subsection{Prolate spheroidal coordinates}
	We model an elongated target by the surface of a three-dimensional prolate spheroid (i.e., an ellipsoid of revolution) with the single major semiaxis $b$ along the $z$ coordinate and two equal minor semiaxes $a<b$,
	
	\begin{equation}
		\Gamma=\left\{\left(x,y,z\right) \in \mathbb{R}^{3}: \frac{x^{2}}{a^{2}}+\frac{y^2}{a^{2}}+\frac{z^2}{b^{2}}=1\right\} ,
	\end{equation}
	surrounded by a concentric prolate spheroid with the single major semiaxis $B$ along the $z$ coordinate and two equal minor semiaxes $A<B$:
	
	\begin{equation}
		\partial \Omega_{0}=\left\{\left(x,y,z\right) \in \mathbb{R}^{3}: \frac{x^{2}}{A^{2}}+\frac{y^2}{A^{2}}+\frac{z^2}{B^{2}}=1\right\}.
	\end{equation}
	
\begin{figure}[!ht]
		\centering
		\subfloat[$\alpha_1 = 0.1$, $\alpha_2=1$ \\$\frac{a}{b}=0.10$, $\frac{A}{B}=0.76$]{\includegraphics[width=0.25\linewidth]{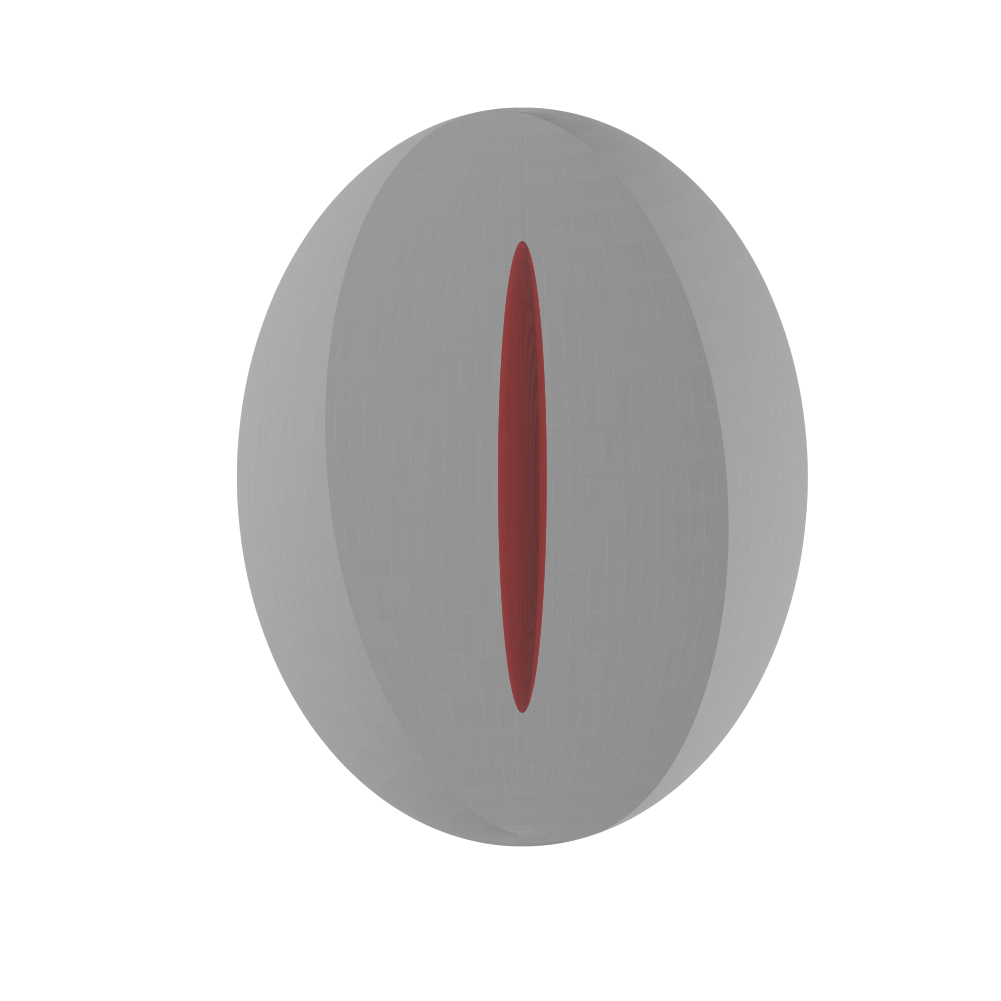}}
		\subfloat[$\alpha_1 = 0.5$, $\alpha_2=1$ \\$\frac{a}{b}=0.46$, $\frac{A}{B}=0.76$]{\includegraphics[width=0.25\linewidth]{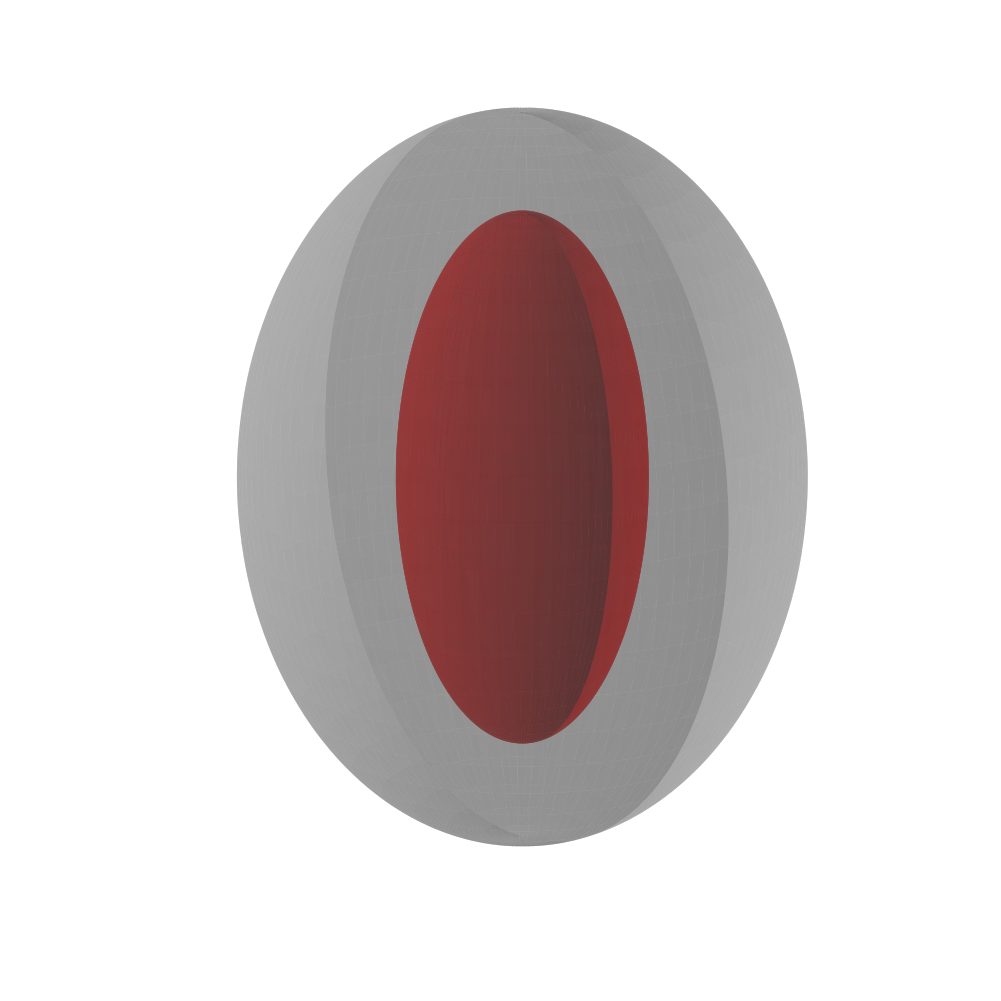}}
		\subfloat[$\alpha_1 = 0.9$, $\alpha_2=1$ \\$\frac{a}{b}=0.72$, $\frac{A}{B}=0.76$]{\includegraphics[width=0.25\linewidth]{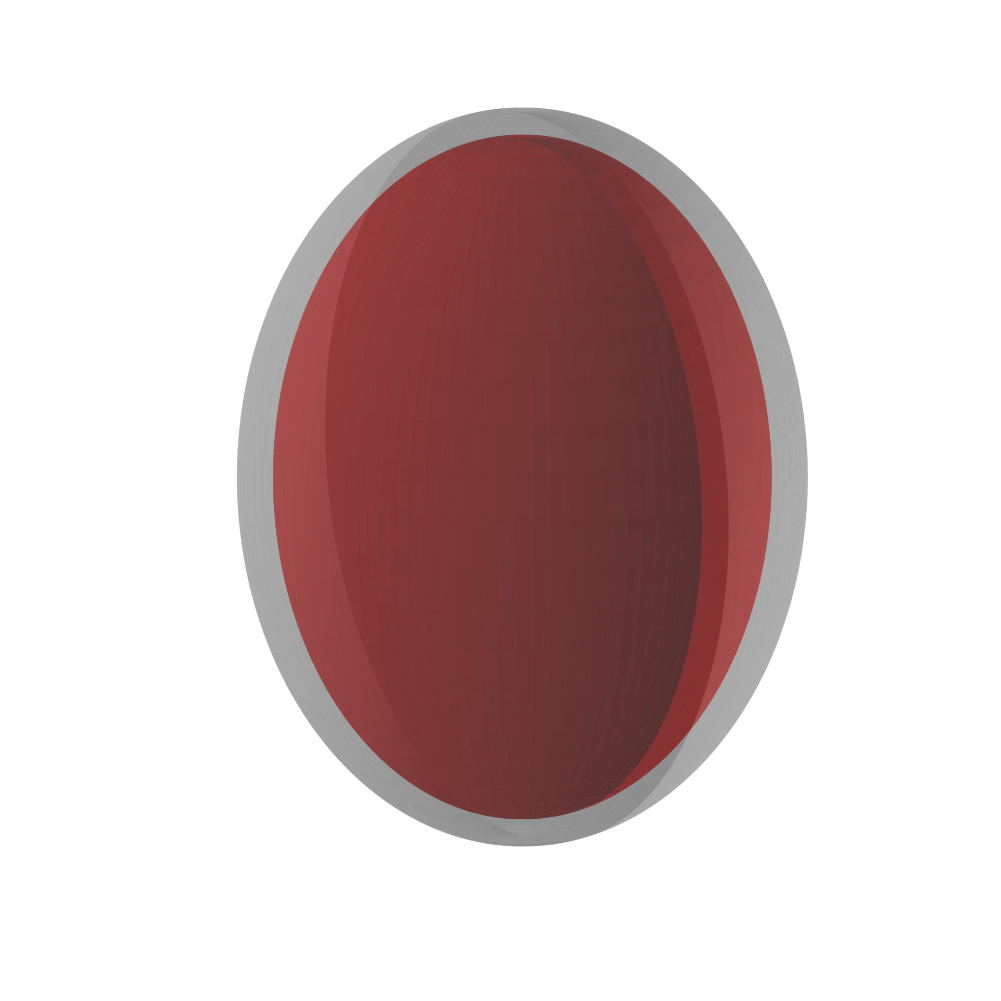}}
		\hfil
		\subfloat[$\alpha_1 = 0.1$, $\alpha_2=2$\\$\frac{a}{b}=0.10$, $\frac{A}{B}=0.96$]{\includegraphics[width=0.25\linewidth]{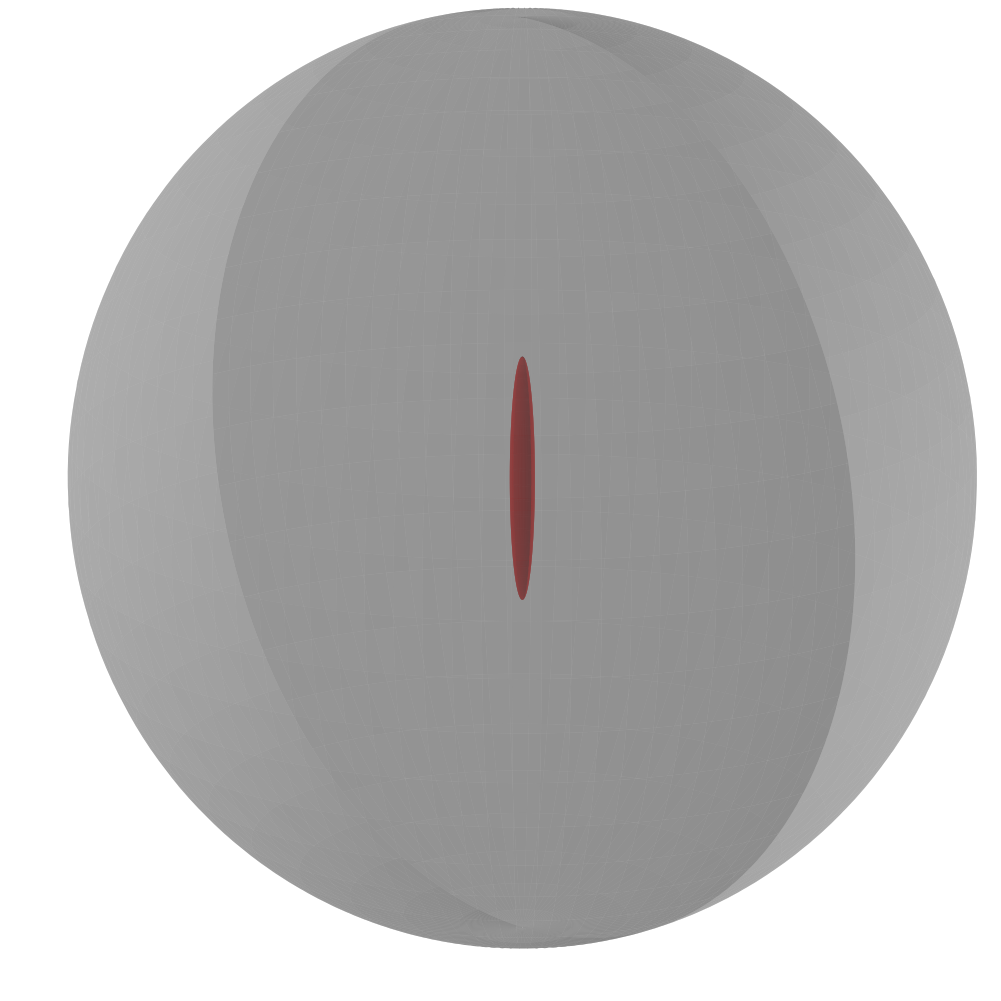}}
		\subfloat[$\alpha_1 = 0.5$, $\alpha_2=2$\\$\frac{a}{b}=0.46$, $\frac{A}{B}=0.96$]{\includegraphics[width=0.25\linewidth]{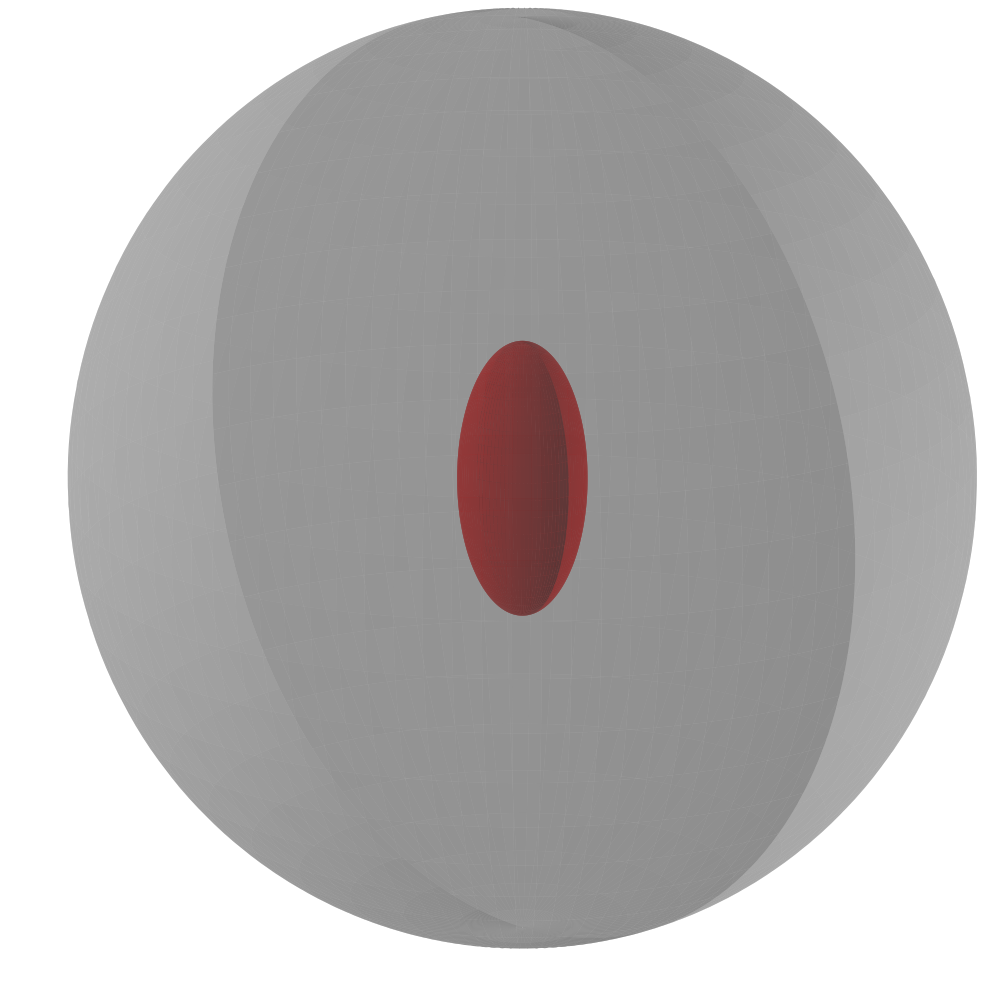}}
		\subfloat[$\alpha_1 = 0.9$, $\alpha_2=2$\\$\frac{a}{b}=0.72$, $\frac{A}{B}=0.96$]{\includegraphics[width=0.25\linewidth]{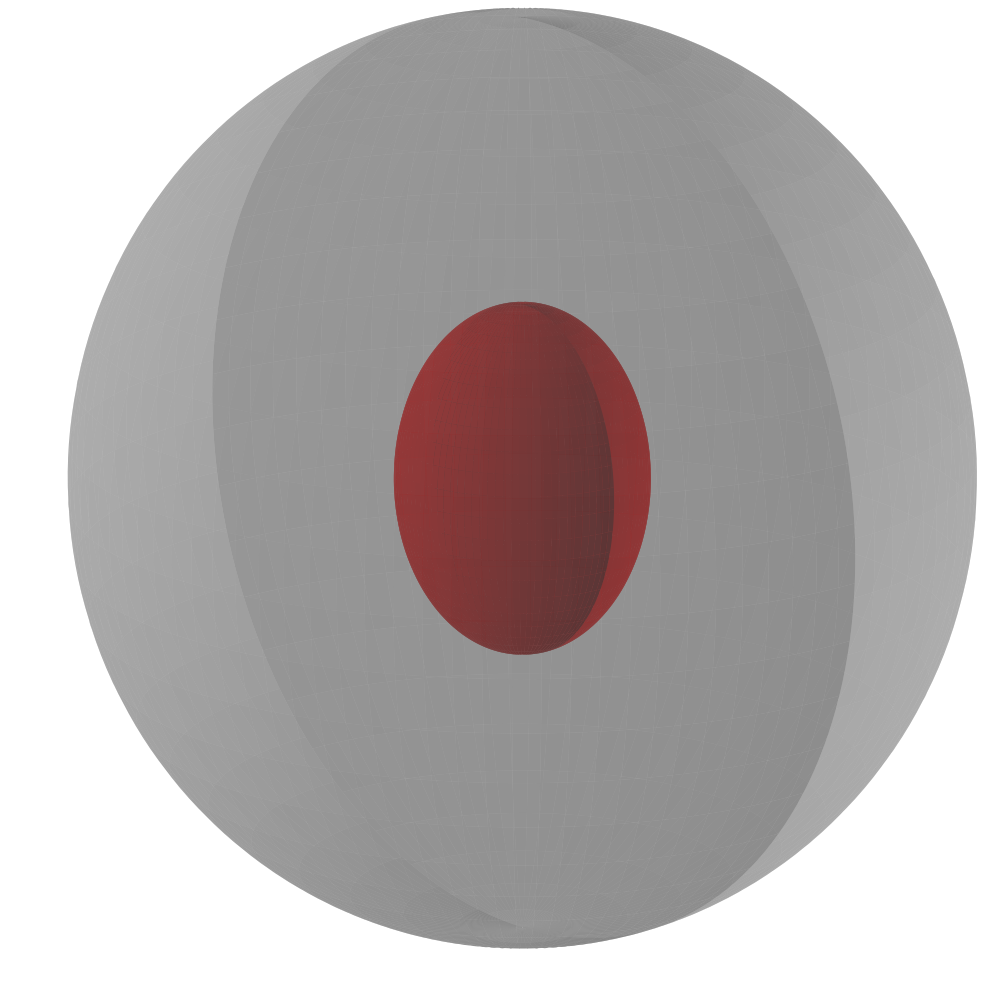}}
		\caption{Examples of concentric prolate spheroids with $c=1$, $\alpha_{1} \in \left\{0.1, 0.5, 0.9 \right\}$ and $\alpha_{2} = \left\{1, 2\right\}$. For the first line the ratio $b/B$ is respectively 0.65, 0.73, 0.93. For the second line the ratio $b/B$ is respectively 0.27, 0.30, 0.38.}
		\label{fig:ellipsoids}
	\end{figure}

	We introduce the  prolate spheroidal coordinates $(\alpha, \theta, \phi)$, that are related to the Cartesian coordinates $(x,y,z)$ as 
	
	\begin{equation}
		\left(\begin{array}{l}
			x \\
			y \\
			z
		\end{array}\right)= c \left(\begin{array}{c}
			\sinh \alpha \sin \theta \cos \phi \\
			\sinh \alpha \sin \theta \sin \phi \\
			\cosh \alpha \cos \theta
		\end{array}\right),
	\end{equation}
	where $0 \leq \alpha<\infty$, $0\leq \theta \leq \pi$, $0 \leq \phi <2 \pi$, and
	
	\begin{equation}
		\left(\begin{array}{c}
			\alpha \\
			\theta \\
			\phi
		\end{array}\right)=\left(\begin{array}{c}
			\cosh ^{-1}\left[\left(r_{+}+r_{-}\right) /\left(2 c\right)\right] \\
			\cos ^{-1}\left[\left(r_{+}-r_{-}\right) /\left(2 c\right)\right] \\
			\tan ^{-1}\left(y / x\right)
		\end{array}\right),
	\end{equation}
	where $r_{\pm} = \sqrt{x^2 + y^2 + (z \pm c)^2}$ are the distances to the two foci located at points $(0,0, \pm c)$ and 
	
	\begin{equation}\label{eq:c}
		c=\sqrt{b^2-a^2} = \sqrt{B^2 - A^2}
	\end{equation} is half of the focal distance. Note that this relation introduces a constraint on the shapes of two spheroids. In this new coordinate system the domain $\Omega$ is defined as 
	\begin{equation}
		\Omega = \left\{ \alpha_{1} < \alpha < \alpha_{2}\text{, } 0 \leq \theta \leq \pi\text{, } 0 \leq \phi < 2\pi \right\},
	\end{equation} 
	where $\alpha_{1} = \tanh^{-1}(a/b)$ determines the target boundary
	\begin{equation}
		\Gamma = \left\{ \alpha = \alpha_{1} \text{, } 0\leq \theta \leq \pi \text{, } 0\leq \phi < 2\pi \right\},
	\end{equation}
	and $\alpha_{2} = \tanh^{-1}(A/B)$ determines the outer reflecting boundary
	\begin{equation}
		\partial \Omega_0 = \left\{ \alpha = \alpha_{2} \text{, } 0\leq\theta\leq\pi \text{, } 0\leq\phi<2\pi \right\}.
	\end{equation}
	In particular, the smallness of the target is determined by the condition
	
	\begin{equation}\label{eq:smallness}
		\frac{b}{B} = \frac{\cosh \alpha_{1}}{\cosh \alpha_{2}} \ll 1.
	\end{equation}
	Note that $a/b = \tanh\alpha_{1}$ and $A/B=\tanh\alpha_2$ characterize the anisotropy of the target $\Gamma$ and of the outer boundary $\partial \Omega_0$, respectively. When the ratio approaches 1 the shape is close to a sphere; in turn, when the ratio approaches 0 the shape is highly anisotropic (elongated). Figure \ref{fig:ellipsoids} illustrates different configurations of the domain $\Omega$ between two concentric spheroids.
	
	 The volume and surface area of prolate spheroids are well known, in particular, the volume of the confining domain $\Omega$ is
	\begin{equation}\label{eq:vol}
		|\Omega| = \frac{4\pi}{3} \left( A^2B -  a^2b\right),
	\end{equation}
	and the surface area of the target $\Gamma$ is
	\begin{equation}
		|\Gamma| = 2\pi a^2 \left(1+ \frac{b}{ae}\sin^{-1}(e)\right), \quad \text{with } e= \sqrt{1- \cfrac{a^2}{b^2}}.
	\end{equation}
	For further derivations, we use the scale factors of the change of coordinates
	\begin{align}
		h_{\alpha} &= c \sqrt{\sinh^2(\alpha) + \sin^2(\theta)}, \label{eq:ha}  \\ 
		h_{\theta} &= c \sqrt{\sinh^2(\alpha) + \sin^2(\theta)}, \label{eq:ht} \\ 
		h_{\phi} &= c \sinh\alpha \sin\theta. \label{eq:hp}
	\end{align}
	
	\subsection{Harmonic measure density}
\begin{figure}[!ht]
		\centering
		\includegraphics[width=0.65\linewidth]{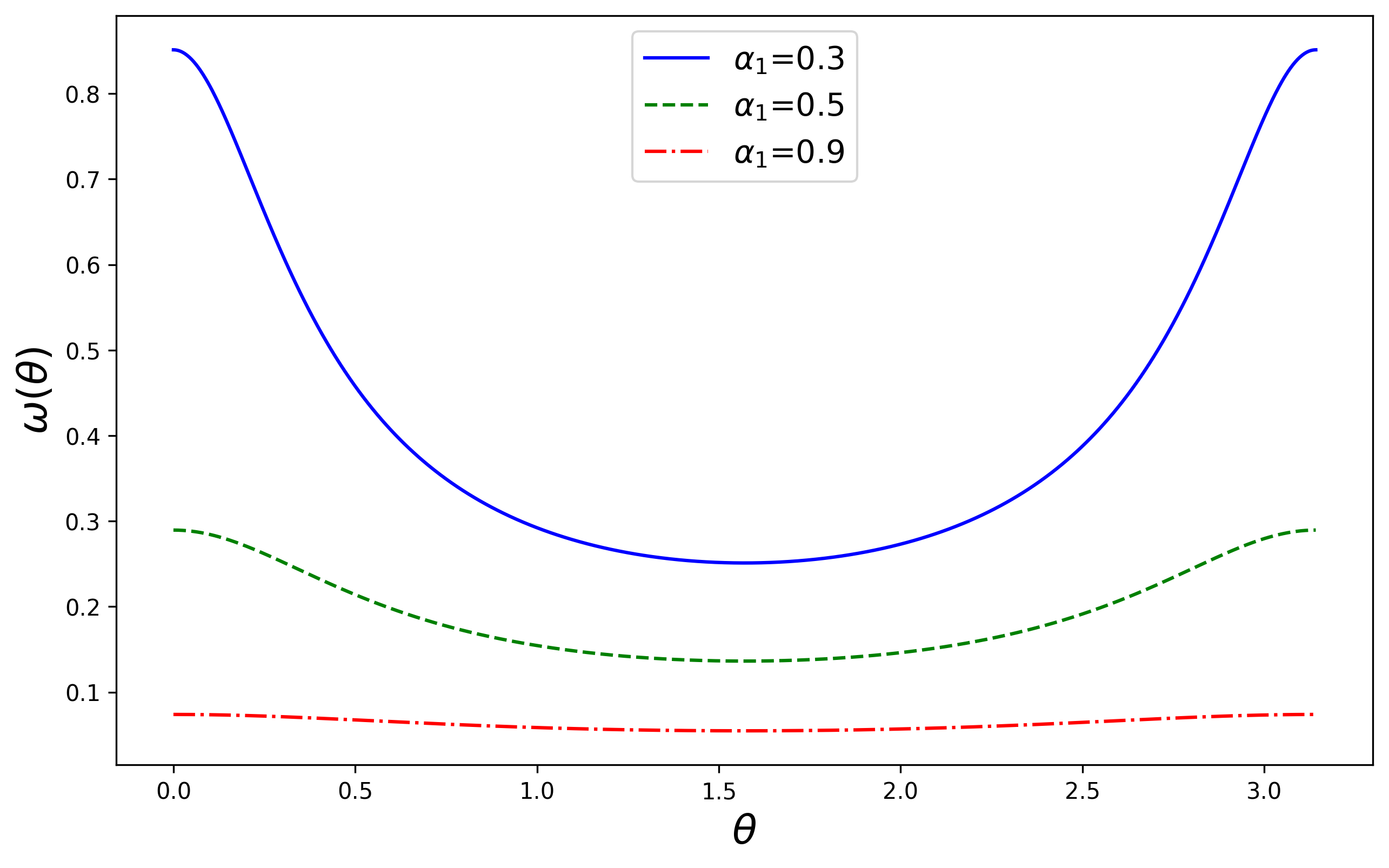}
		\caption{The volume-averaged harmonic measure density from Eq. (\ref{eq:mharmofin}) as a function of the angle $\theta$, for prolate spheroids with $c=1$ and $\alpha_2=2$.}
		\label{fig:omega_2.15}
	\end{figure}

	The derivation of the volume-averaged harmonic measure density is based on an explicit representation of the Green's function in prolate spheroidal coordinates (see \ref{app:pro}). Taking the normal derivative and integrating over the starting point according to its definition (\ref{eq:defw}), we get after a lengthy computation
	
	\begin{equation}\label{eq:mharmofin}
		\omega(\boldsymbol{x}) = \frac{1}{4\pi c \sinh\alpha_{1}h_{\alpha_{1}}(\theta)} \left(1-\frac{8\pi c^3\sinh^2(\alpha_1)}{15|\Omega|} I_2 P_2(\cos\theta)\right),
	\end{equation}
	where 
	\begin{equation}
		\label{eq:I2}
		\begin{aligned}[b]
			I_{2} &=-\frac{5}{6} \frac{P_{2}^{\prime}\left(\cosh \alpha_{1}\right) Q_{2}^{\prime}\left(\cosh \alpha_{2}\right)-Q_{2}^{\prime}\left(\cosh \alpha_{1}\right) P_{2}^{\prime}\left(\cosh \alpha_{2}\right)}{P_{2}\left(\cosh \alpha_{1}\right) Q_{2}^{\prime}\left(\cosh \alpha_{2}\right)-Q_{2}\left(\cosh \alpha_{1}\right) P_{2}^{\prime}\left(\cosh \alpha_{2}\right)},
		\end{aligned}
	\end{equation}
	and $P_n(x)$ and $Q_n(x)$ are the Legendre functions of the first and second kind, and prime denotes the derivative with respect to the argument. In particular, $P_2(x) = \frac{1}{2}(3x^2-1)$ and $Q_2(x) = \frac{3x^2-1}{4} \ln\left(\frac{x+1}{x-1}\right) - \frac{3x}{2}$. The first arrival  position $\boldsymbol{x}$ is fully characterized by the angle $\theta$ (the axial symmetry implies that $\omega(\boldsymbol{x})$ does not depend on $\phi$).
	Figure \ref{fig:omega_2.15} illustrates the behavior of the volume-averaged harmonic measure density  in Eq. (\ref{eq:mharmofin}). As $\alpha_1$ gets smaller (i.e. the target becomes more anisotropic), the volume-averaged harmonic measure density exhibits stronger variations with $\theta$, with two maxima on the extremities of the prolate spheroid which are the most exposed to Brownian motion.
	
	%
	Let us inspect Eq. (\ref{eq:mharmofin}) in more details. One sees that there are two effects of the angle $\theta$ onto the function $\omega(\boldsymbol{x})$: via a multiplicative factor proportional to $1/h_{\alpha_1}(\theta)$ and via an additive term proportional to $P_2(\cos\theta)$. The first effect is a consequence of a non-linear parameterization of the target surface by curvilinear coordinates $\theta$ and $\phi$. In fact, the surface element on the target surface is 
	\begin{equation}\label{eq:dx}
		d\boldsymbol{x} = h_{\theta}(\theta)h_{\phi}(\theta)d\theta d\phi = c\sinh\alpha_1h_{\alpha_1} d\xi d\phi,
	\end{equation}
	where $\xi=\cos\theta \in (-1,1)$ and we used Eqs. (\ref{eq:ha}--\ref{eq:hp}). The high values of $\omega(\theta)$ at the extremities, which appeared due to the factor $1/h_{\alpha_1}(\theta)$ in Eq. (\ref{eq:mharmofin}), are attenuated by the low density of points there due to the factor $h_{\alpha_1}(\theta)$ in Eq. (\ref{eq:dx}). As these two factors cancel each other, it is more natural to look directly at the probability of the first arrival in a vicinity of point $\boldsymbol{x}$, 
	\begin{equation}
		\omega(\boldsymbol{x}) d\boldsymbol{x} = \hat{\omega}(\xi, \phi) d\xi d\phi,
	\end{equation}
	where 
	\begin{equation}\label{eq:hat_omega}
		\hat{\omega}(\xi, \phi) = c\sinh\alpha_{1}h_{\alpha_{1}} \omega(\boldsymbol{x}) = \frac{1-\gamma P_2(\cos\theta)}{4\pi},
	\end{equation}
	with 
	\begin{equation}\label{eq:gamma}
		\gamma = \frac{2 I_{2}}{5} \frac{\sinh ^{2}(\alpha_{1})}{\sinh^{2}(\alpha_{2}) \cosh\alpha_{2}-\sinh^{2}(\alpha_{1}) \cosh \alpha_{1}}.
	\end{equation}

	One can easily check that 
	\begin{equation}
		\nonumber \int_{\Gamma} \omega(\boldsymbol{x}) d\boldsymbol{x} = \int_{-1}^{1} \int_{0}^{2\pi} \hat{\omega}(\xi, \phi) d\xi d\phi = 1.
	\end{equation}
	One concludes that the first effect, which was responsible for large variations of $\omega(\theta)$ in Figure \ref{fig:omega_2.15}, is removed  when looking at $\omega(\boldsymbol{x})d\boldsymbol{x}$, or equivalently at $\hat{\omega}(\xi, \phi)$. In the following, we focus on the second effect which is intrinsic and still present in Eq. (\ref{eq:hat_omega}).
	
	Equation (\ref{eq:mharmofin}) or, equivalently (\ref{eq:hat_omega}), presents the main result of this section. It shows how the volume-averaged harmonic measure density depends on the location of the arrival point $\boldsymbol{x}$ through $\xi = \cos \theta$. The anisotropy of the target makes $\hat{\omega}(\xi, \phi)$ non-uniform, which is controlled by the parameter $\gamma$ given by Eq. (\ref{eq:gamma}). We checked numerically that $\gamma>0$ for all $\alpha_1<\alpha_2$.
	When the target is small, the parameter $\gamma$ is expected to be small as well. According to Eq. (\ref{eq:smallness}), the relative smallness of the target can be ensured by setting $\alpha_2 \to \infty$. In this limit, the shape of the target is fixed, while the outer boundary goes to infinity.
	
\begin{figure}[!ht]
		\centering
		\includegraphics[width=0.65\linewidth]{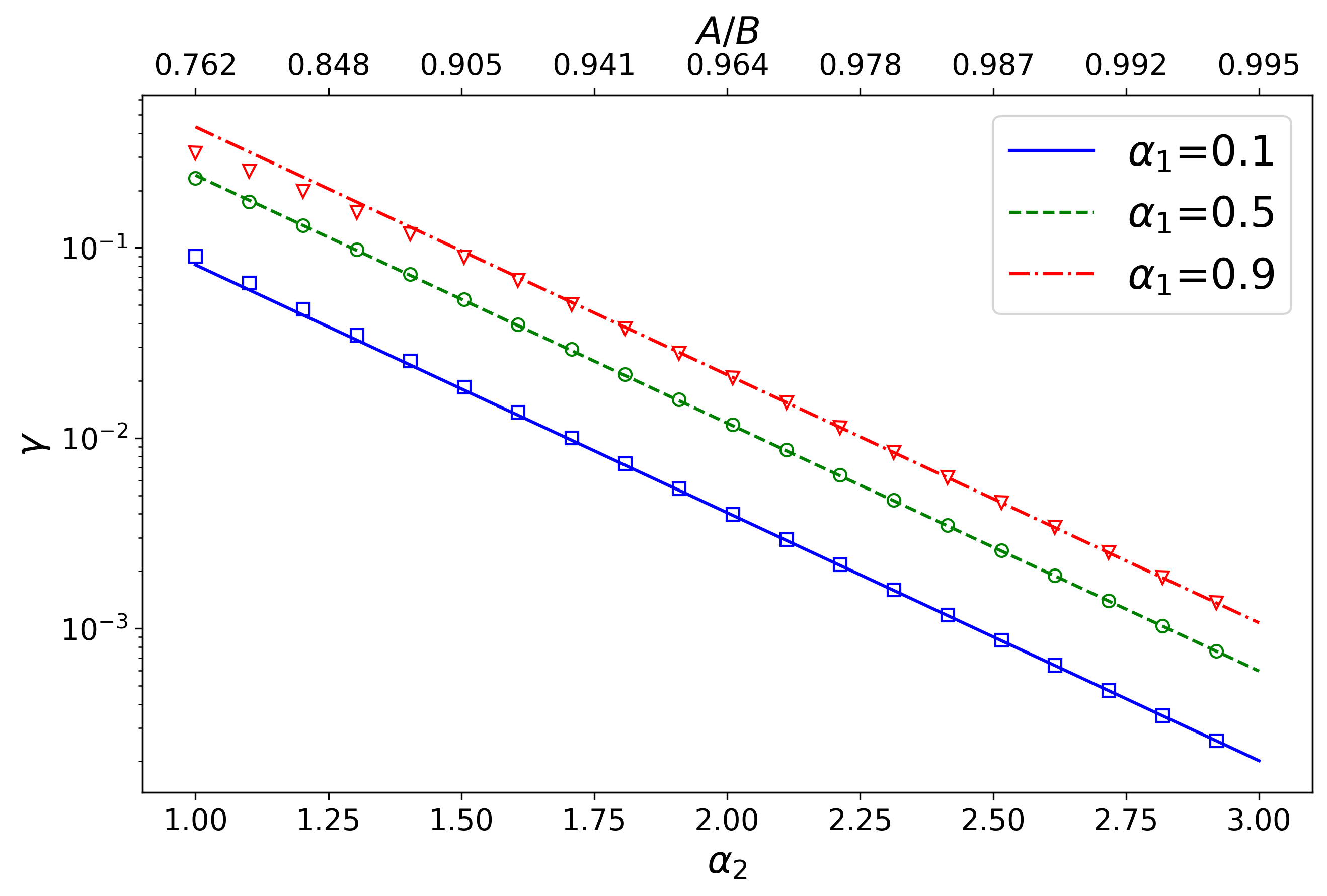}
		\caption{Semi log plot of the coefficient $\gamma$ in Eq. (\ref{eq:gamma}) (in symbols) as a function of $\alpha_{2}$ for three values of $\alpha_{1}$, with $c=1$. Lines present the asymptotic behavior  (\ref{eq:gamma_asymp}) of $\gamma$. Note that the values $A/B$ equivalent to $\alpha_2$ are shown on the top.}
		\label{fig:gammavsao}
	\end{figure}

	Using the asymptotic behavior of Legendre functions we get
	\begin{equation}\label{eq:gamma_asymp}
		\gamma \approx -e^{-3 \alpha_{2}} \frac{8 \sinh ^{2}(\alpha_{1})}{3} \frac{Q_{2}^{\prime}\left(\cosh \alpha_{1}\right)}{Q_{2}\left(\cosh \alpha_{1}\right)} \quad (\alpha_2 \gg 1).
	\end{equation}
	The symbol $\approx$ denotes the asymptotic behavior of $\gamma$ when $\alpha_2$ goes to infinity; however, it also emphasizes that the left-hand side is close to the right-hand side when $\alpha_2$ is large enough.
	One sees that $\gamma$ vanishes exponentially fast with $\alpha_{2}$. Since $B/c = \cosh\alpha_{2}\approx e^{\alpha_2}/2$,
	one also gets 
	$\gamma \approx B^{-3} \propto 1/|\Omega|$
	in this limit.  As the volume of the confining domain grows, $\hat{\omega}(\xi, \phi)$ becomes almost uniform (constant), i.e. all target points are (almost) equally accessible to Brownian motion.
	Figure \ref{fig:gammavsao} shows the dependence of the coefficient $\gamma$ on the size of the domain; one sees that $\gamma$ vanishes exponentially when the outer boundary gets larger. \\
	
\begin{figure}[!ht]
		\centering
		\includegraphics[width=0.65\linewidth]{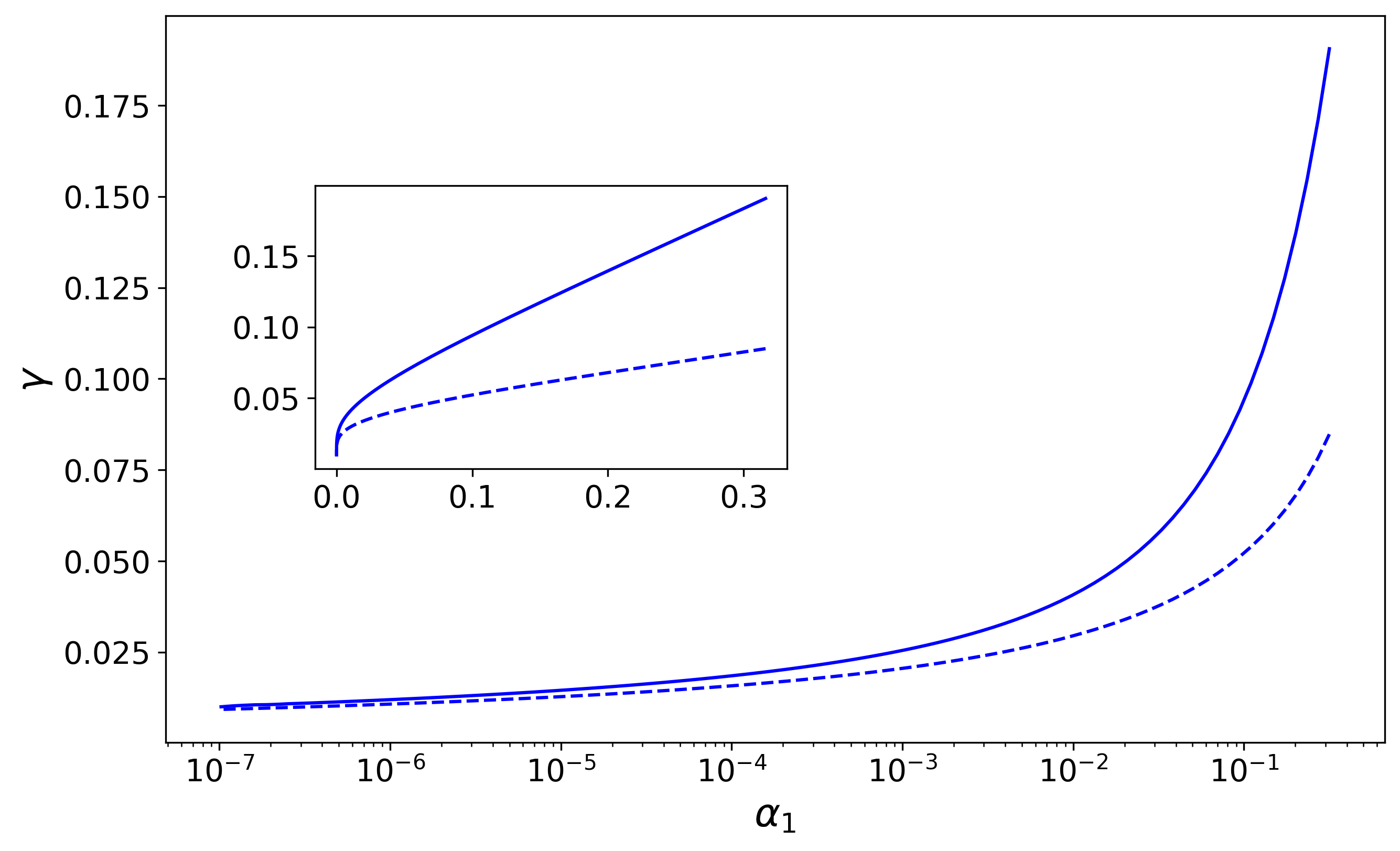}
		\caption{Semi log plot of the coefficient $\gamma$ in Eq. (\ref{eq:gamma}) (in solid line) as a function of $\alpha_{1}$ for $\alpha_2=1$, with $c=1$. Dashed-line presents the asymptotic behavior  (\ref{eq:gamma_asymp2}). The inset shows the same plot with the linear horizontal axis. }
		\label{fig:gammavsa1}
	\end{figure}

	We also briefly discuss the limit $\alpha_1 \to 0$ when the outer boundary is fixed, while the target tends to a segment of length $2c$, which is never hit by diffusion. 
	Using the asymptotic behavior of Legendre functions we get
	\begin{equation}\label{eq:gamma_asymp2}
		\gamma \approx \frac{1}{3\sinh^2(\alpha_2)\cosh\alpha_2 \ln(2/\alpha_1)} \quad (\alpha_1 \ll 1).
	\end{equation}
	One sees that  $\gamma$ exhibits a very slow decay as $1/\ln(2/\alpha_1)$, as illustrared in Figure \ref{fig:gammavsa1}.
	This result suggests that the inaccessibility of the target also makes its points almost equally (in)accessible.
	
	\color{black}
	\subsection{Mean first-passage time}
	
	The expression (\ref{eq:G}) of the Green's function $G(\boldsymbol{x}, \boldsymbol{x}_0)$ allows us to derive the mean first-passage time $T(\boldsymbol{x}_0)$ from Eq. (\ref{eq:defT}) as 
	
	\begin{equation}
		T(\boldsymbol{x}_0) =T_0(\alpha) - T_2(\alpha) P_2(\cos\theta),
	\end{equation}
	where $\boldsymbol{x}_0= (\alpha, \theta, \phi)$ is now the starting point,
	\begin{align}\label{eq:T0}
		T_0(\alpha) = \frac{c^{2}}{3D}\Bigl[&\frac{\cosh^{2}(\alpha_{1})-\cosh^2(\alpha)}{2} \\
		\nonumber &+ \sinh^2(\alpha_{2}) \cosh\alpha_{2}\left(Q_{0}\left(\cosh \alpha_{1}\right)-Q_{0}(\cosh \alpha)\right)\Bigr],\\
		T_2(\alpha) =\frac{c^2}{9D} \biggl(1 &+
		\frac{P'_2(\cosh\alpha_2) Q_2(\cosh\alpha)
			-Q'_2(\cosh\alpha_2) P_2(\cosh\alpha)}
		{P'_2(\cosh\alpha_2) Q_2(\cosh\alpha_1)
			-Q'_2(\cosh\alpha_2) P_2(\cosh\alpha_1)}
		\biggr).
	\end{align}
	In the limit $a \to b$ and $A \to B$, one should retrieve the mean first-passage time to a perfectly reactive spherical target of radius $\rho = a$ surrounded by a reflecting sphere of radius $R=A$ \cite{grebenkov2018strong} 
	
	\begin{equation}\label{eq:tsphere}
		T(\boldsymbol{x}_0) = \frac{(|\boldsymbol{x}_0|-\rho)\left(2 R^{3}-\rho |
			\boldsymbol{x}_0|(|\boldsymbol{x}_0|+\rho)\right)}{6D |\boldsymbol{x}_0| \rho}.
	\end{equation}
	
	For illustration purposes, we choose the reflecting spheroidal boundary $\partial \Omega_{0}$ to be close to a sphere of radius $1$, so that only the target exhibits anisotropy. For this purpose, we set $A = 0.99$ and $B = 1.01$ and thus $\alpha_{2}=\tanh^{-1}(A/B)\approx 2.30$.
	
\begin{figure*}[!ht]
		\centering
		\subfloat[]{\label{fig:tc_a}\includegraphics[width=0.325\linewidth]{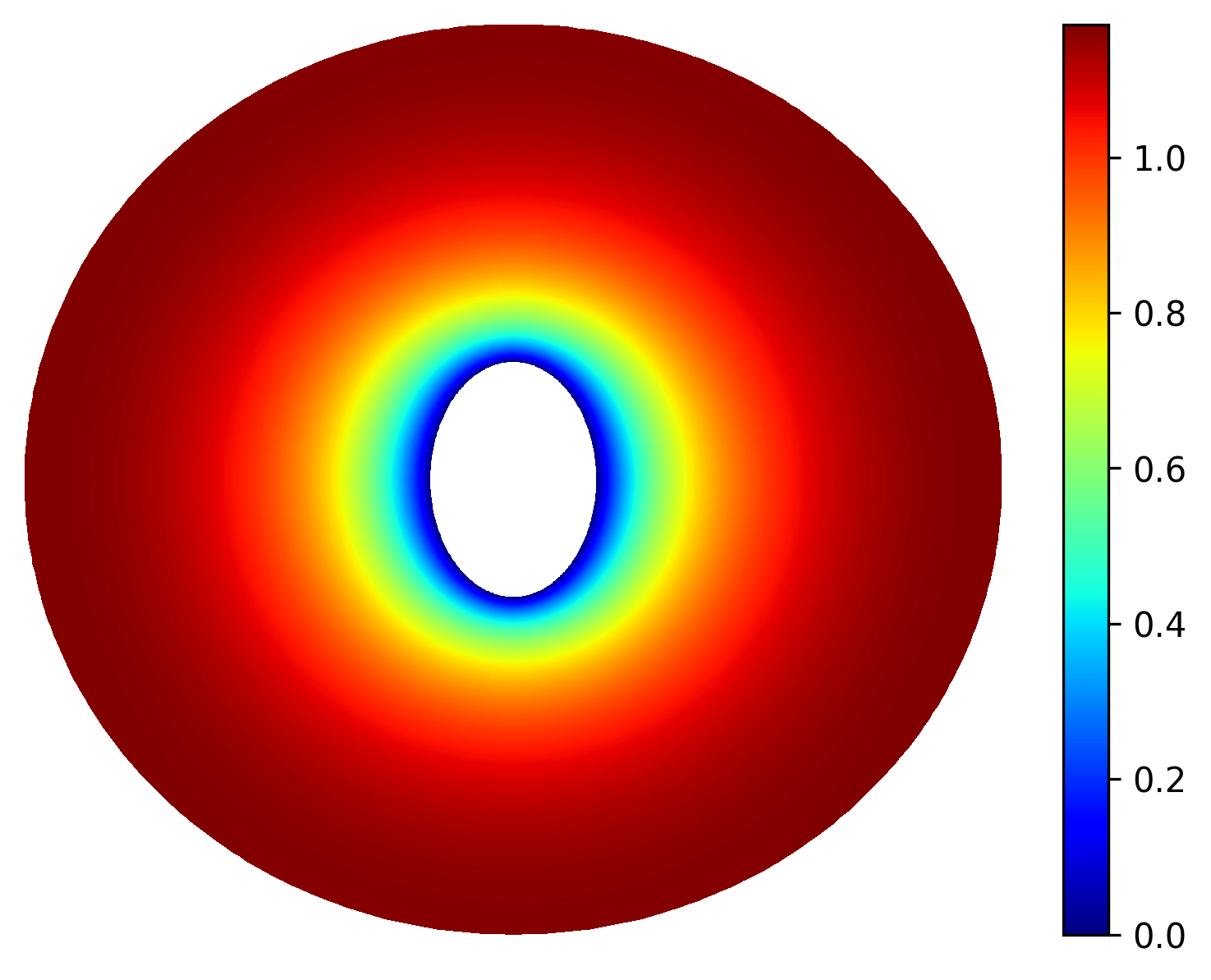}}
		\subfloat[]{\label{fig:tc_b}\includegraphics[width=0.325\linewidth]{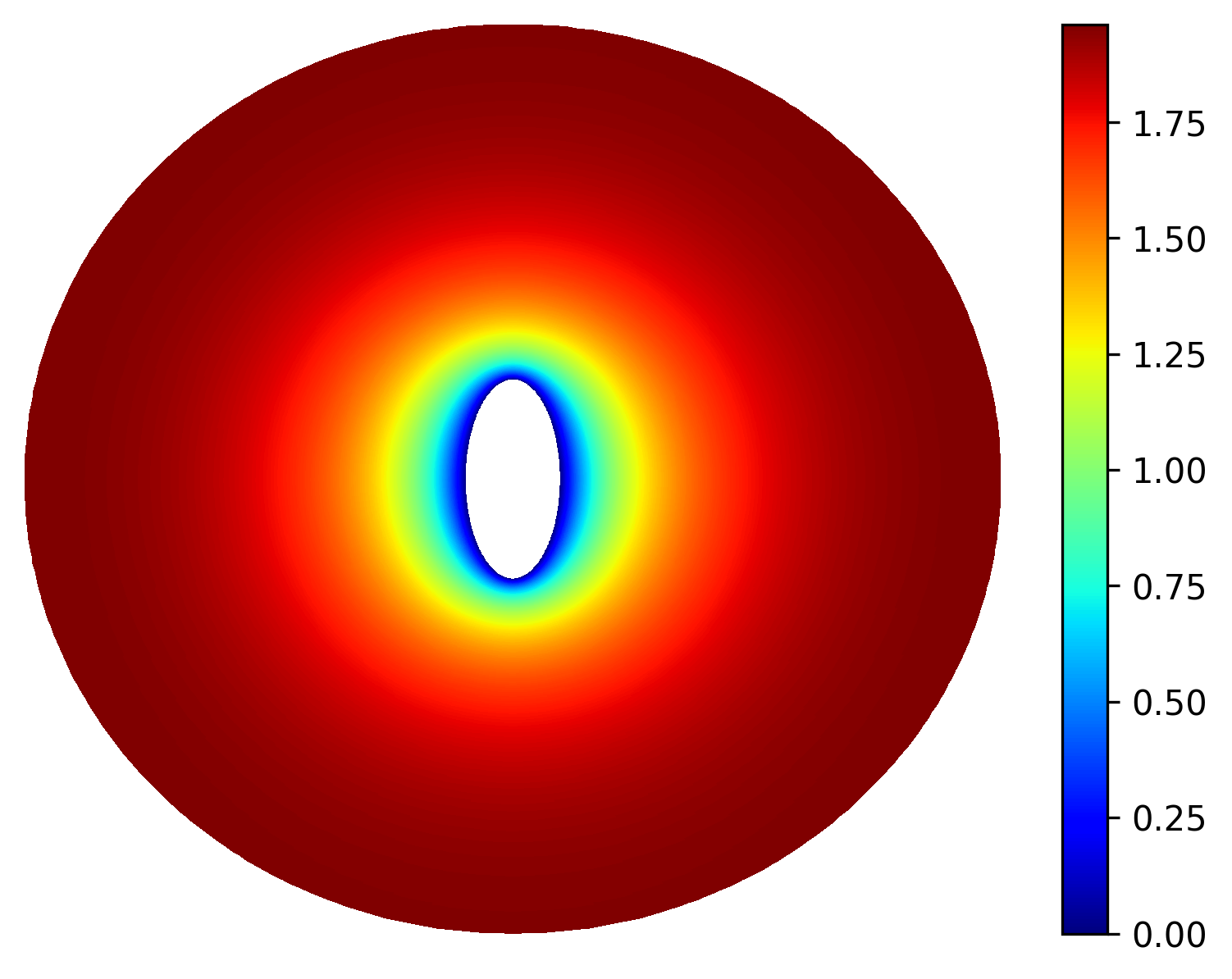}}
		\hfil
		\subfloat[]{\label{fig:tc_c}\includegraphics[width=0.325\linewidth]{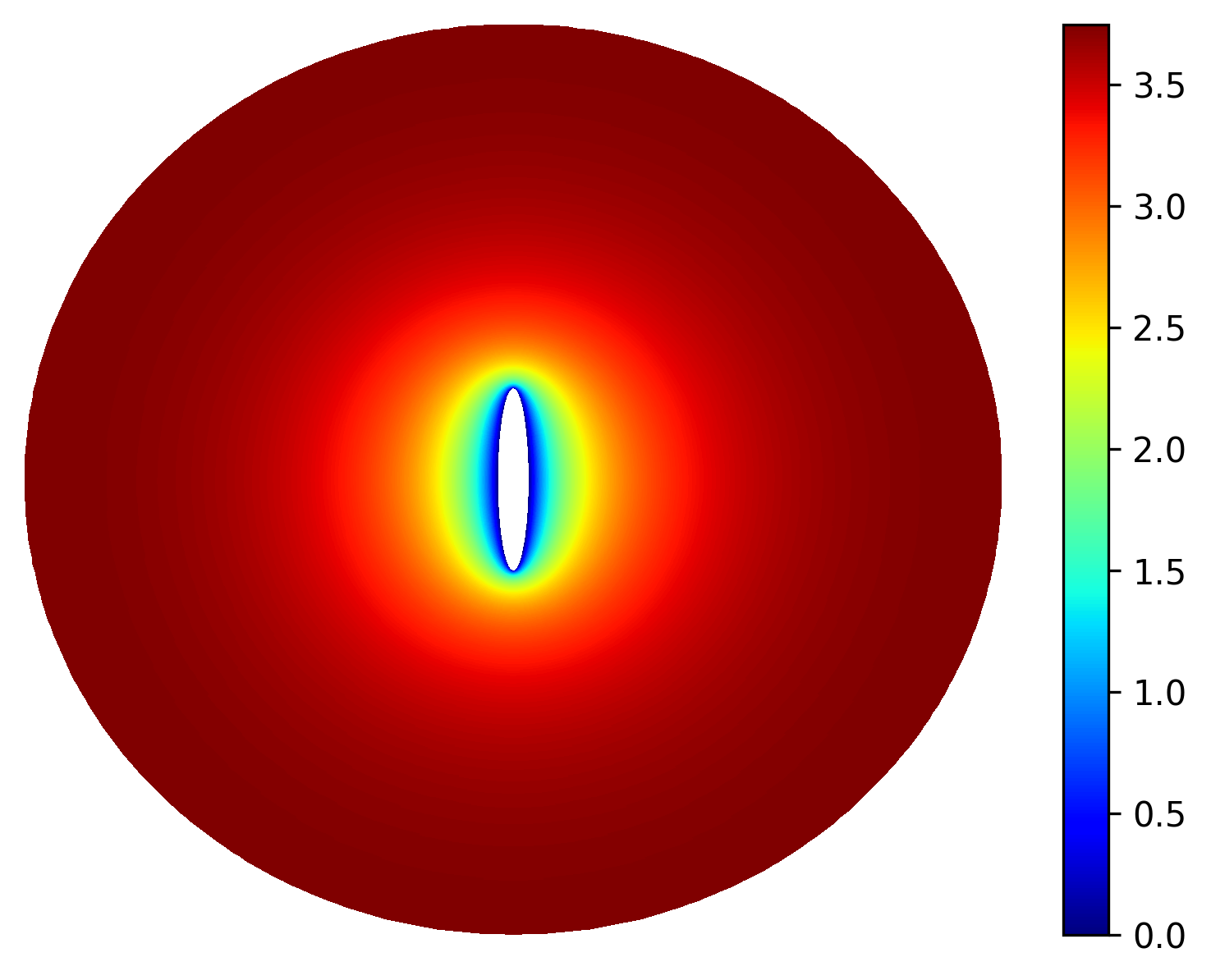}}
		\subfloat[]{\label{fig:tc_d}\includegraphics[width=0.325\linewidth]{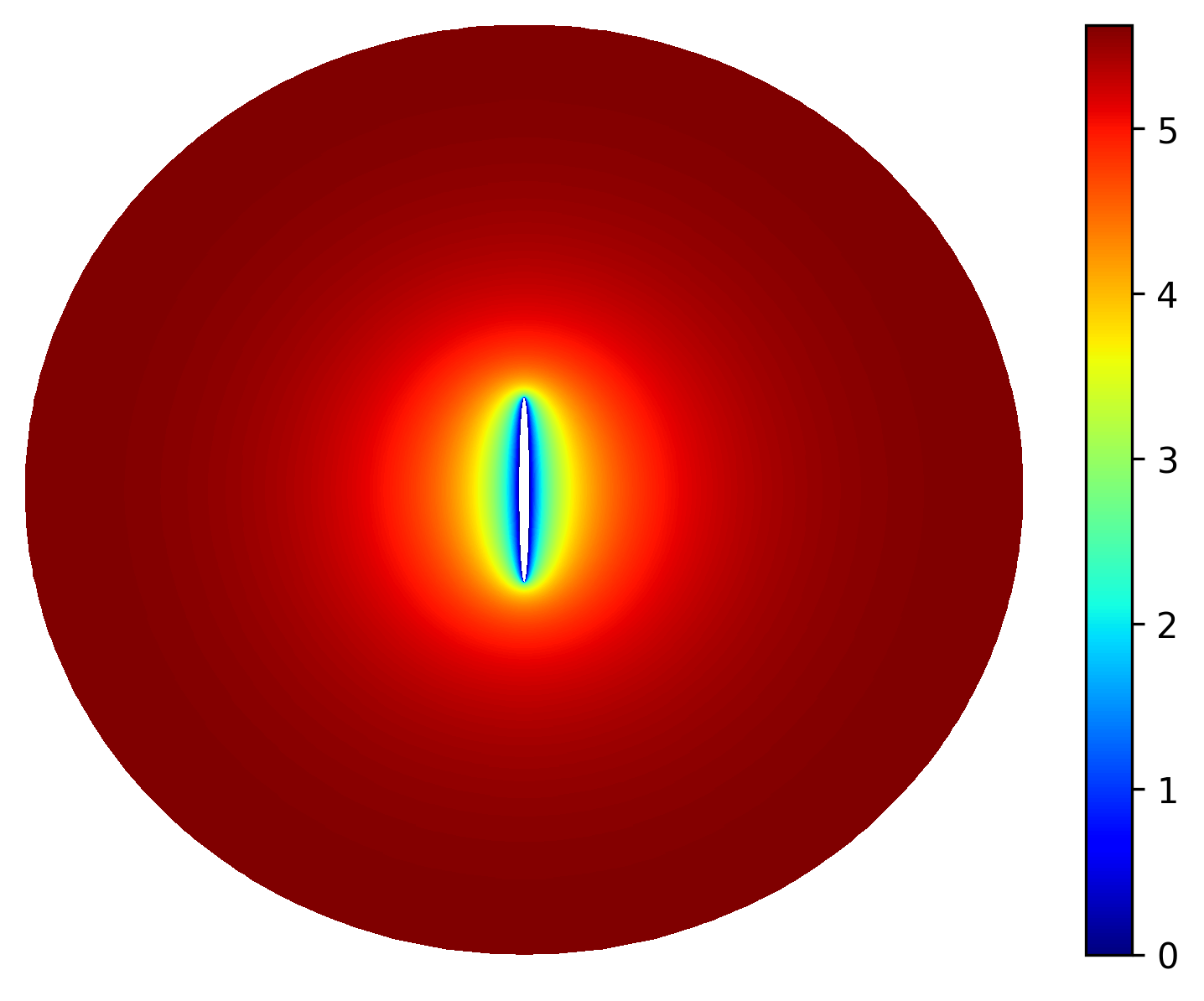}}
		\caption{Cross-section along the $z$-axis of the mean first-passage time $T(\boldsymbol{x}_0)$ to the target as a function of the starting point $\boldsymbol{x}_0$ of the particle for several spheroidal targets (in white) with semiaxes: 
			(a) $a \approx 0.17$ et $b \approx 0.26$, (b) $a \approx 0.10$ et $b \approx 0.22$, (c) $a \approx 0.03$ et $b \approx 0.20$, (d) $a \approx 0.01$ et $b \approx 0.20$ surrounded by a ``roundish'' concentric prolate spheroid  with semiaxis $A=0.99$ and $B=1.01$. We set $D=1$.}
		\label{fig:tm_coupe}
	\end{figure*}

	Figure \ref{fig:tm_coupe} illustrates the mean first-passage time to the target as a function of the starting point of the particle. It shows that for a ``roundish'' target the mean first-passage time increases symmetrically in all directions as one moves away from the target and one retrieves the classical behavior (\ref{eq:tsphere}) for a spherical target. In turn, for anisotropic targets, the mean first passage-time increases with distorsion depending on the shape of the target.
	
	When the starting point $\boldsymbol{x}_0$ is located far away from the target, the anisotropy effect is greatly reduced. To illustrate this effect, we set the starting point on the outer boundary, i.e., $\alpha = \alpha_2$.
	In this case one can easily check that $T_0(\alpha_2)$ exhibits the asymptotic behavior 
	
	\begin{equation}\label{eq:T_as1}
		T_0(\alpha_{2}) \approx \frac{c^2}{24D} e^{3\alpha_{2}} Q_0(\cosh \alpha_{1}) \quad (\alpha_2\gg1),
	\end{equation} 
	that is to say, $T_0(\alpha_2)$ exponentially grows with $\alpha_2$ as the domain increases while 
	\begin{equation}
		T_2(\alpha_{2}) \approx \cfrac{c^2}{9D} \quad (\alpha_{2}\gg1).
	\end{equation} 
	In other words, for a particle diffusing from the outer boundary the dependence on the starting point (i.e., the dependence on $\theta$) is insignificant, i.e.
	
	\begin{equation}
		T(\boldsymbol{x}_0) \approx T_0(\alpha_2) \quad (\alpha_{2}\gg1).
	\end{equation}	
	
\begin{figure}[!ht]
		\centering
		\includegraphics[width=0.65\linewidth]{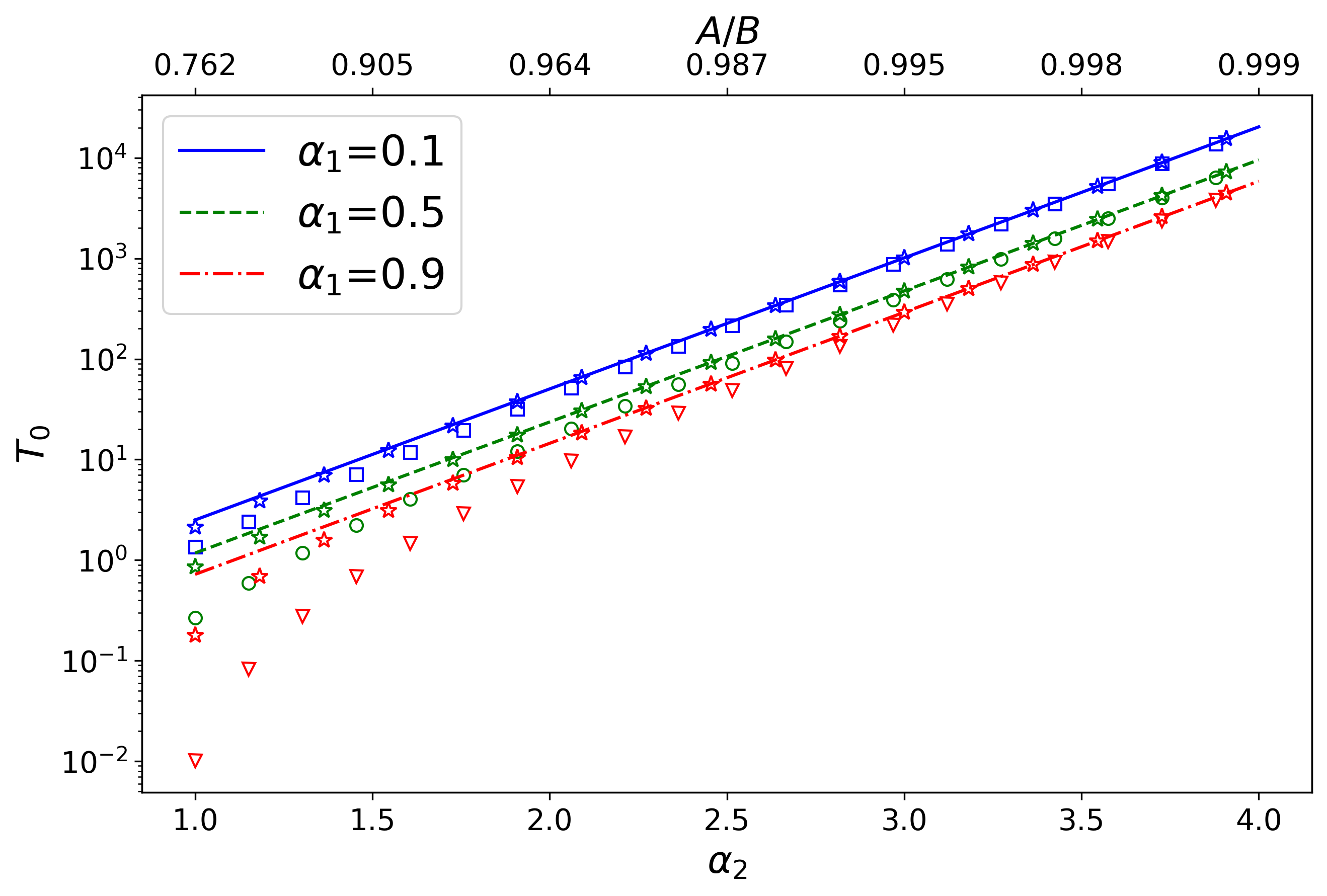}
		\caption{Semi log plot of $T_0(\alpha_{2})$ from Eq. (\ref{eq:T0}) (in symbols) as a function of $\alpha_{2}$ for three values of $\alpha_{1}$, with $c=1$ and $D=1$. Lines present the asymptotic behavior  (\ref{eq:T_as1}) of $T(\boldsymbol{x}_0)$, while stars present the asymptotic relation (\ref{eq:T_as2}). Note that the values of $A/B$ equivalent to $\alpha_2$ are shown on top.}
		\label{fig:T0}
	\end{figure}

	Moreover, using the expression (\ref{eq:vol}) of the volume $|\Omega|$ and the capacity $C$ of a prolate spheroid in three dimensions \cite{landau2013electrodynamics}
	\begin{equation}
		C = \cfrac{8\pi c}{\ln\left(\frac{1+c/b}{1-c/b}\right)},
	\end{equation}
	we easily check the expected capacitance approximation \cite{cheviakov2011optimizing, kolokolnikov2005optimizing, maz1984asymptotic, ward1993strong}:
	\begin{equation}\label{eq:T_as2}
		\overline{T} = \frac{1}{|\Omega|}\int_{\Omega}T(\boldsymbol{x}_0)d\boldsymbol{x}_0 \approx \frac{|\Omega|}{D C} \approx \frac{c^2}{24D} e^{3\alpha_{2}} Q_0(\cosh \alpha_{1}) \quad (\alpha_2\gg1).
	\end{equation}
	Indeed, integrating $T(\boldsymbol{x}_0)$ over the volume we get 
	
	\begin{equation}
		\overline{T} = \frac{\pi c^5 I}{D|\Omega|},
	\end{equation}
	with 
	\begin{align*}
		I =& -\frac{4\cosh\alpha_{1}}{135} \big[3\cosh^4\alpha_1 -5\cosh^2\alpha_1 -2\big] + \frac{4}{9} \cosh^2\alpha_1\cosh\alpha_2\sinh^2\alpha_2 \\
		&- \frac{8}{135} \cosh\alpha_2 \big[6\cosh \alpha_2 - 5\cosh^2\alpha_2 +1 \big] \\
		&+\frac{4}{9} \sinh^4\alpha_2\cosh^2\alpha_2 \big[Q_0(\cosh\alpha_1) - Q_0(\cosh\alpha_2)\big] \\
		&+ \frac{4}{45} \sinh^2\alpha_1 \frac{P_2'(\cosh\alpha_2)Q_2'(\cosh\alpha_1) - Q_2'(\cosh\alpha_2)P_2'(\cosh\alpha_1)}{P_2'(\cosh\alpha_2)Q_2(\cosh\alpha_1) - Q_2'(\cosh\alpha_2)P_2(\cosh\alpha_1)}.
	\end{align*}
	Figure \ref{fig:T0} illustrates these asymptotic behaviors for a particle diffusing from the outer boundary. On this semi log plot, one sees the expected exponential growth of $T_0(\alpha_2)$ when the size of domain increases, and the relevance of the asymptotic relations (\ref{eq:T_as1}) and (\ref{eq:T_as2}).
	
\begin{figure}[!ht]
		\centering
		\includegraphics[width=0.65\linewidth]{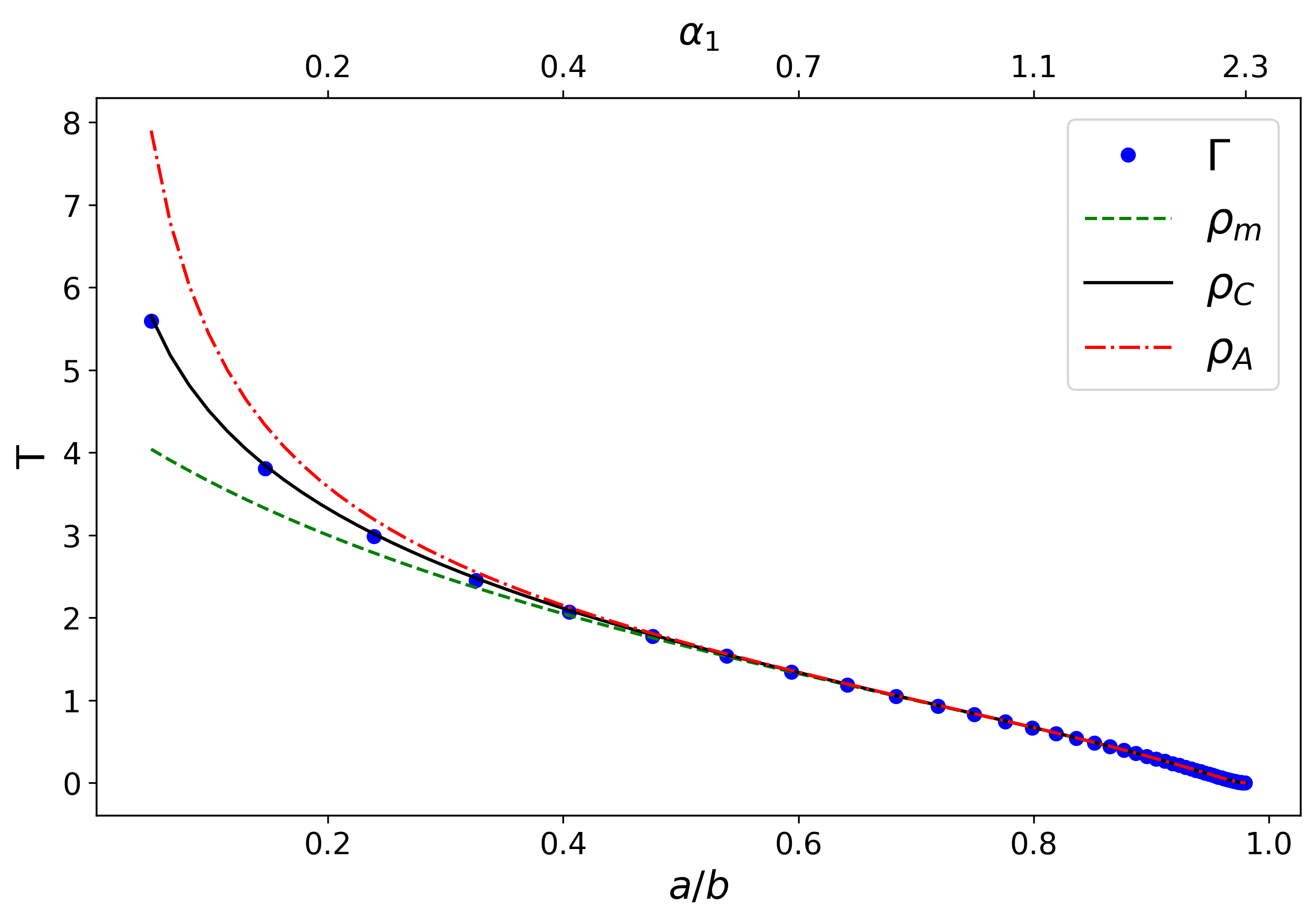}
		\caption{Mean first-passage time to a prolate spheroidal target $\Gamma$ (shown by filled circles) as a function of its aspect ratio $a/b$ for a particle diffusing from the outer boundary -- a concentric spheroid with semiaxis $A=0.99$ and $B=1.01$ which implies $\alpha_{2}=\tanh^{-1}(A/B)\approx 2.30$ and $c \approx 0.04$ according to Eq. (\ref{eq:c}). For comparison, three curves shown by lines present the mean first-passage time to an equivalent spherical target with three choices $\rho_m$, $\rho_C$, $\rho_A$ of the effective radius.}
		\label{fig:mfpt_pro}
	\end{figure}

	We compare the mean first-passage time of a particle diffusing from the outer boundary to a prolate spheroidal target and that to  an ``equivalent'' spherical target.
	It is important to underline that the choice of the  criterion of equivalence between a sphere and a spheroid is central. For example, given a prolate spheroid of two minor semiaxes $a$ and major semiaxis $b$, one can choose a sphere whose radius is the mean of the  semiaxes of the spheroid:
	
	\begin{equation}
		\rho_m = (2a+b)/3.
	\end{equation}
	With this configuration the particle always reaches the sphere faster (compare blue circles and dashed green line on Figure \ref{fig:mfpt_pro}).  \\
	
	One can consider other criteria of ``equivalence''. Since the harmonic capacity plays a major role in diffusion-reaction processes \cite{chaigneau2022first, maz1984asymptotic}, we can set the radius of the sphere $\rho_C$ such that the spherical target and the spheroidal one have the same harmonic capacity:
	
	\begin{equation}
		\rho_C = \frac{2c}{\ln\left(\frac{1+c/b}{1-c/b}\right)}.
	\end{equation}
	Here, it appears that the mean first-passage times are very close to each other (compare blue circles and solid black line on Figure \ref{fig:mfpt_pro}), even when the target is small. 
	
	One can also think about the equivalence in terms of optimization. For example, how to minimize the mean first-passage time to the target given a certain amount of reactants uniformly distributed on the target boundary (i.e. given a surface area $|\Gamma|$). In our study (limited to spheres and spheroids) we set the radius of the spherical target $\rho_A$ such that the surface areas of both targets are equal:
	
	\begin{equation}
		\rho_A = \sqrt{\frac{|\Gamma|}{4\pi}}.
	\end{equation}
	This time, it is curious to remark that the mean first-passage time to the anisotropic target is smaller than the mean first-passage time to the spherical target (compare blue circles and dash-dotted red line on Figure \ref{fig:mfpt_pro}), meaning that for a given target surface a prolate spheroid presents a better ``trapping ability''. This difference is enhanced even more when the reactive surface is reduced and the target anisotropy increases. 
	
	In all three cases, the mean first-passage time shown on Figure \ref{fig:mfpt_pro} vanishes as $a/b$ approaches one. This is a consequence  of the constraint (\ref{eq:c}). In fact, as $A$ and $B$ are fixed, $c$ is also fixed. To make $a/b$ close to one, one should increase both $a$ and $b$ (under the constraint $b^2-a^2=c^2$), i.e. enlarge the target that makes it closer to the starting point on the outer boundary and thus diminishes the mean first-passage time. To eliminate the growth of the target, one has to modify the shape of the outer boundary, as we did in Figure \ref{fig:T0}.
	\section{Flattened targets}\label{sec:oblate}
	In this section, we consider the domain $\Omega$ between biaxial concentric oblate spheroids in three dimensions and replicate the earlier analysis for such domains.
	
	\subsection{Oblate spheroidal coordinates}
	
\begin{figure}[!ht]
		\centering
		\subfloat[$\alpha_1=0.1$, $\alpha_2=1$\\$\frac{a}{b}\approx 0.01$, $\frac{A}{B}\approx 0.76$]{\includegraphics[width=0.25\linewidth]{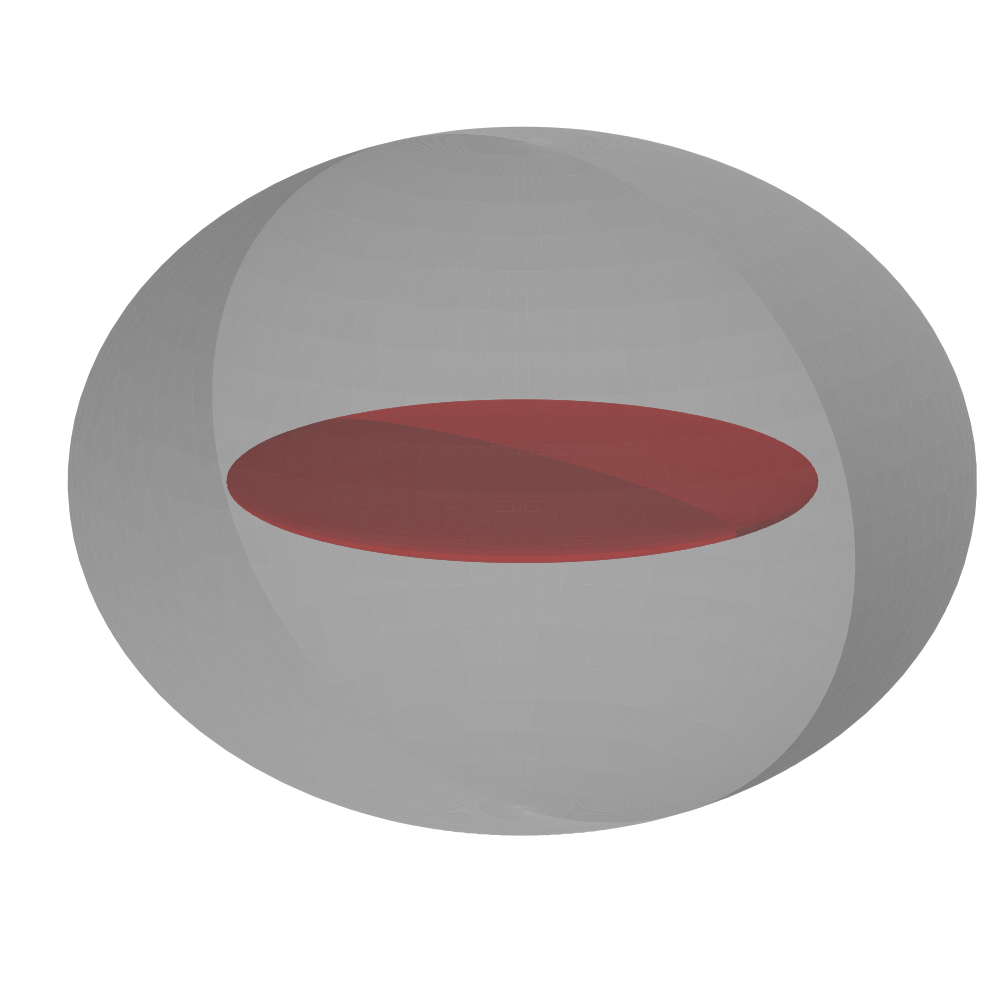}}
		\subfloat[$\alpha_1=0.5$, $\alpha_2=1$\\$\frac{a}{b} \approx 0.46$, $\frac{A}{B}\approx 0.76$]{\includegraphics[width=0.25\linewidth]{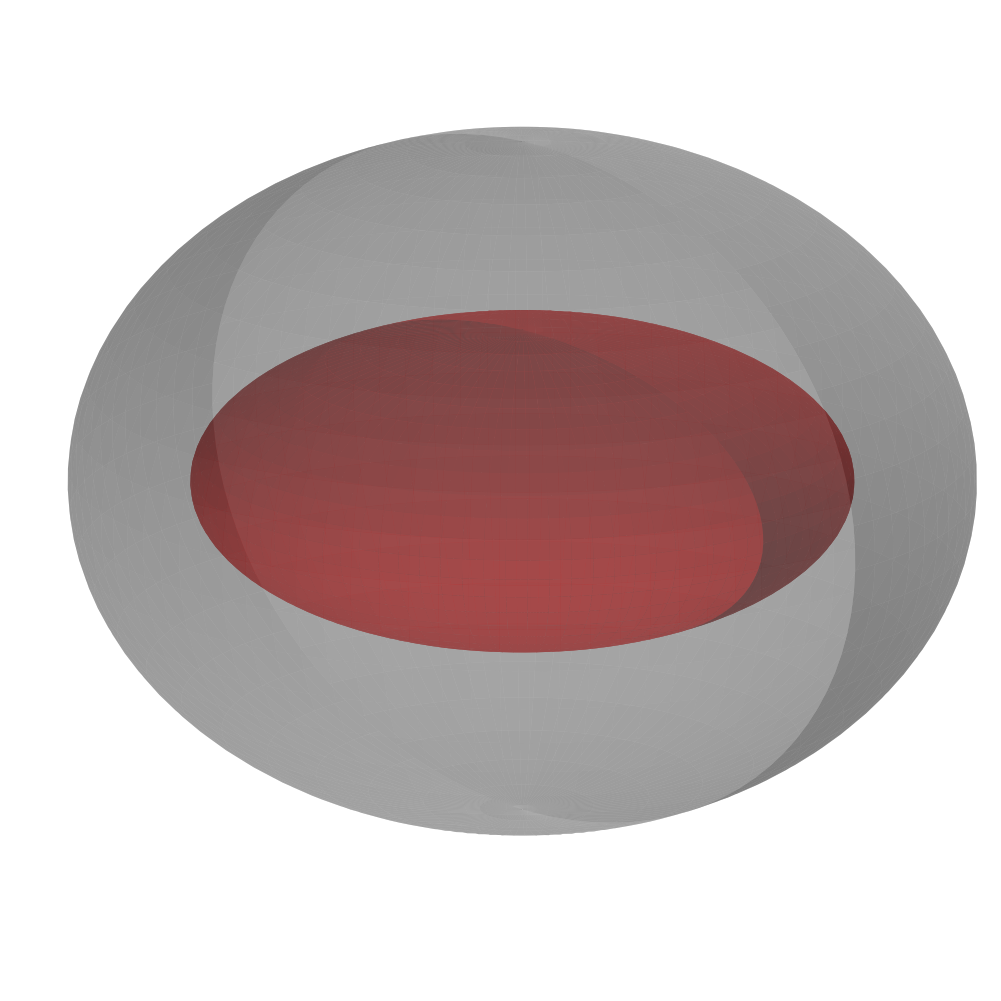}}
		\subfloat[$\alpha_1=0.9$, $\alpha_2=1$\\$\frac{a}{b} \approx 0.72$, $\frac{A}{B}\approx 0.76$]{\includegraphics[width=0.25\linewidth]{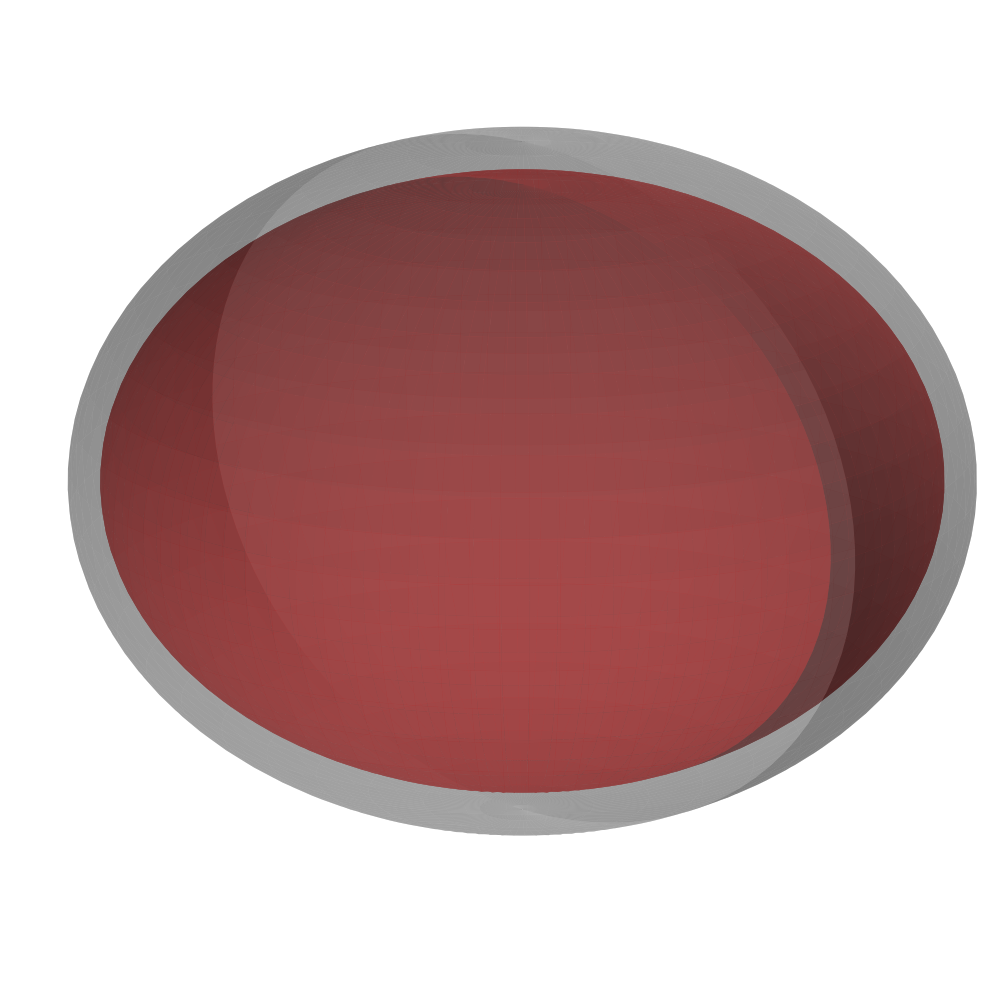}}
		\hfil
		\subfloat[$\alpha_1=0.1$, $\alpha_2=2$\\$\frac{a}{b}\approx 0.01$, $\frac{A}{B}\approx 0.96$]{\includegraphics[width=0.25\linewidth]{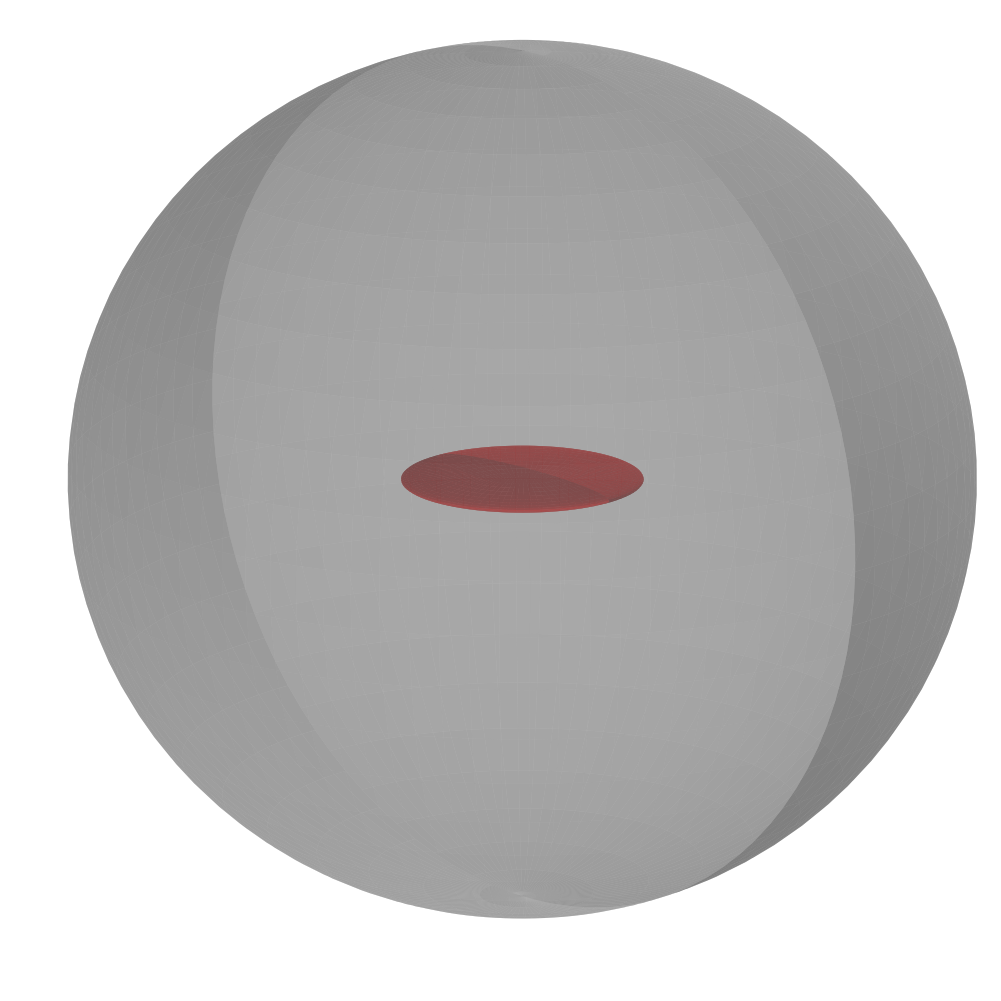}}
		\subfloat[$\alpha_1=0.5$, $\alpha_2=2$\\$\frac{a}{b} \approx 0.46$, $\frac{A}{B}\approx 0.96$]{\includegraphics[width=0.25\linewidth]{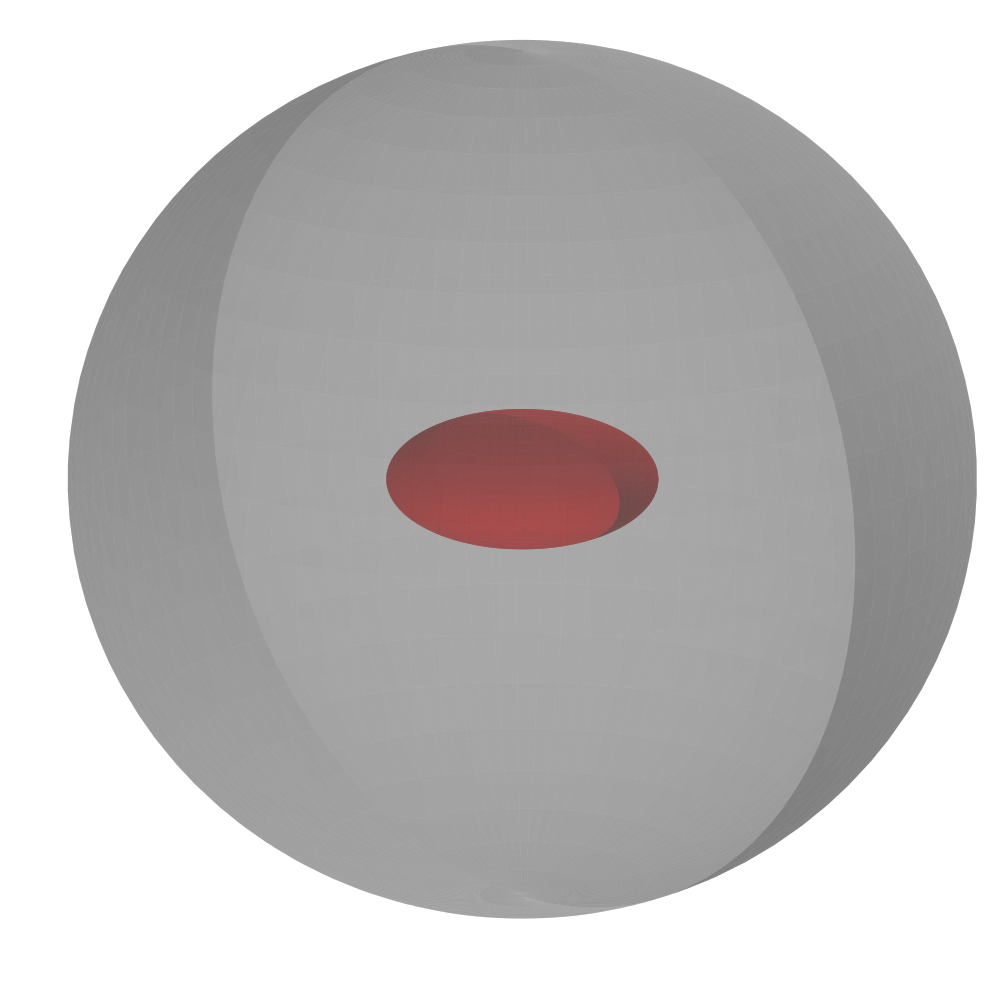}}
		\subfloat[$\alpha_1=0.9$, $\alpha_2=2$\\$\frac{a}{b} \approx 0.72$, $\frac{A}{B}\approx 0.96$]{\includegraphics[width=0.25\linewidth]{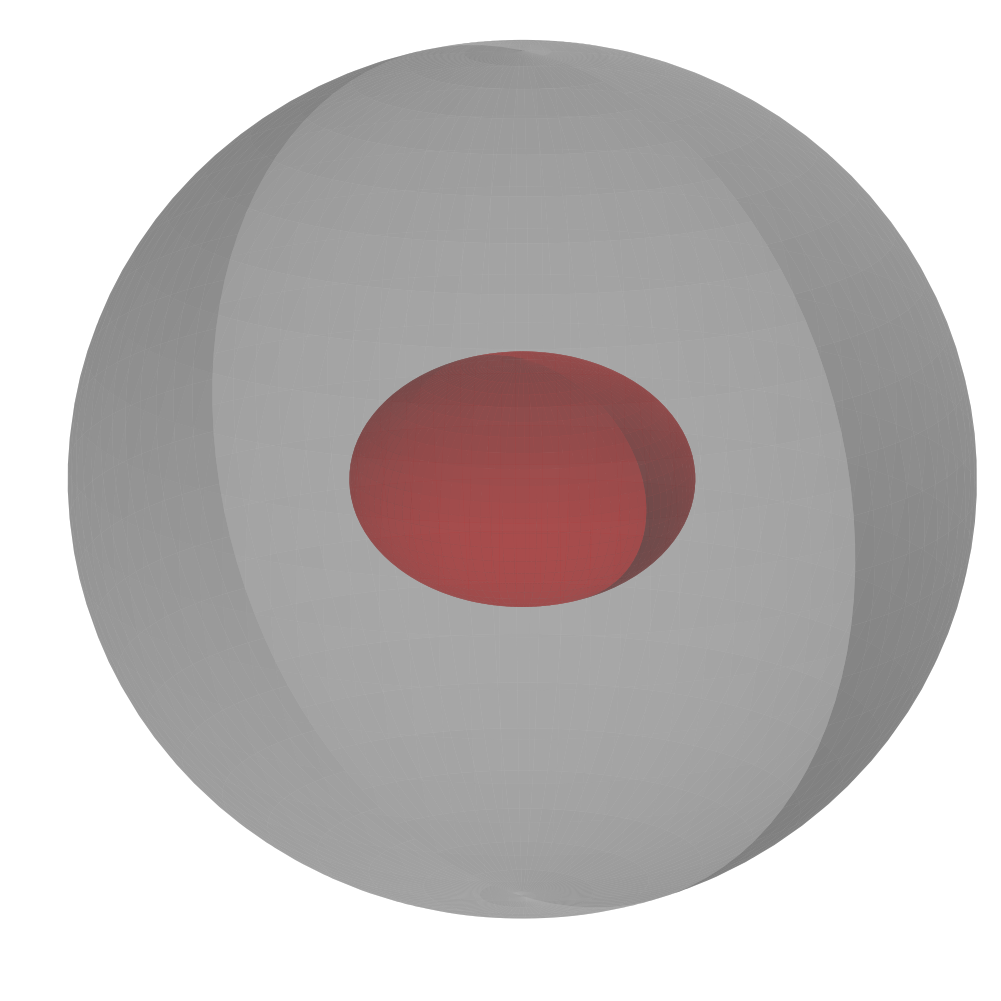}}
		\caption{Examples of concentric oblate spheroids with $c=1$, $\alpha_{1} \in \left\{0.1, 0.5, 0.9 \right\}$ and $\alpha_{2} = \left\{1, 2\right\}$. For the first line the ratio $b/B$ is respectively 0.65, 0.73, 0.93. For the second line the ratio $b/B$ is respectively 0.27, 0.30, 0.38.}
		\label{fig:ellipsoids_obl}
	\end{figure}

	We model a  flattened target by the surface of a three-dimensional oblate spheroid (i.e., an ellipsoid of revolution) with the single minor semiaxis $a$ along the $z$ coordinate and two equal major semiaxes $b>a$:
	
	\begin{equation}
		\Gamma=\left\{\left(x,y,z\right) \in \mathbb{R}^{3}: \frac{x^{2}}{b^{2}}+\frac{y^2}{b^{2}}+\frac{z^2}{a^{2}}=1\right\},
	\end{equation}
	surrounded by an oblate spheroid with the single minor semiaxis $A$ along the $z$ coordinate and two equal major semiaxes $B>A$:
	
	\begin{equation}
		\partial \Omega_{0}=\left\{\left(x,y,z\right) \in \mathbb{R}^{3}: \frac{x^{2}}{B^{2}}+\frac{y^2}{B^{2}}+\frac{z^2}{A^{2}}=1\right\}.
	\end{equation}
	We introduce the  oblate spheroidal coordinates $(\alpha, \theta, \phi)$ that are related to the Cartesian coordinates $(x,y,z)$ as 
	
	\begin{equation}
		\left(\begin{array}{l}
			x \\
			y \\
			z
		\end{array}\right)= c \left(\begin{array}{c}
			\cosh \alpha \cos \theta \cos \phi \\
			\cosh \alpha \cos \theta \sin \phi \\
			\sinh \alpha \sin \theta
		\end{array}\right),
	\end{equation}
	where $0\leq\alpha\leq\infty \text{, } -\pi/2\leq \theta \leq \pi/2 \text{, } 0\leq \phi < 2 \pi$, and
	
	\begin{equation}
		\left(\begin{array}{c}
			\alpha \\
			\theta \\
			\phi
		\end{array}\right)=\left(\begin{array}{c}
			\cosh ^{-1}\left[\left(r_{+}+r_{-}\right) /\left(2 c\right)\right] \\
			\operatorname{sign}(z) \cos ^{-1}\left[\left(r_{+}-r_{-}\right) /\left(2 c\right)\right] \\
			\tan ^{-1}\left(y / x\right)
		\end{array}\right),
	\end{equation}
	where $r_{\pm} = \sqrt{(\sqrt{x^2+y^2}\pm c)^2 + z^2}$ are the distances to the two foci located at points $(\pm c,0, 0)$ and
	\begin{equation}\label{eq:c2}
		c=\sqrt{b^2-a^2} = \sqrt{B^2-A^2}
	\end{equation}
	is half of the focal distance. \\
	In this new coordinate system the domain $\Omega$ is defined as 
	\begin{equation}
		\Omega = \left\{ \alpha_{1} < \alpha <\alpha_{2}, -\pi/2\leq\theta\leq\pi/2, 0 \leq\phi<2\pi \right\},
	\end{equation} 
	where $\alpha_{1} = \tanh^{-1}\left(\frac{a}{b}\right)$ determines the target boundary 
	\begin{equation}
		\Gamma = \left\{ \alpha = \alpha_{1} \text{, } -\pi/2\leq \theta \leq \pi/2 \text{, } 0\leq \phi< 2\pi \right\},
	\end{equation}
	and $\alpha_{2}=\tanh^{-1}\left(\frac{A}{B}\right)$ determines the outer reflecting boundary
	\begin{equation}
		\partial \Omega_0 = \left\{ \alpha = \alpha_{2} \text{, } -\pi/2 \leq \theta \leq \pi/2 \text{, } 0 \leq \phi < 2\pi \right\}.
	\end{equation}
	As previously, $\alpha_1$ and $\alpha_2$ determine the anisotropy of the target and of the outer boundary, respectively. 	Figure \ref{fig:ellipsoids_obl} illustrates different configurations of the domain $\Omega$ between two concentric oblate spheroids.
	
 	The volume and surface area of oblate spheroids are well known, in particular, the volume of the confining domain $\Omega$ is
	\begin{equation}\label{eq:vol2}
		|\Omega| = \frac{4\pi}{3} \left( AB^2 -  ab^2\right),
	\end{equation}
	and the surface area of the target $\Gamma$ is
	\begin{equation}\label{eq:areag}
		|\Gamma| = 2\pi \left(b^2+ \frac{a^2}{e}\tan^{-1}(e)\right), \quad \text{with } e= \sqrt{1- \cfrac{a^2}{b^2}}.
	\end{equation}
	For further derivations, the scale factors of the change of coordinates are
	\begin{align}
		h_{\alpha} &= c \sqrt{\sinh^2(\alpha) + \sin^2(\theta)}, \\
		h_{\theta} &= c \sqrt{\sinh^2(\alpha) + \sin^2(\theta)}, \\
		h_{\phi} &= c \cosh\alpha \cos\theta.
	\end{align}

	\subsection{Harmonic measure density}
	
	Using the Green's function from \ref{app:obl}, we derive the volume-averaged harmonic density in oblate spheroidal coordinates as 
	
	\begin{equation}
		\begin{aligned}\label{eq:omega_obl}
			\omega(\boldsymbol{x}) &= \frac{1}{4\pi c \cosh\alpha_{1} h_{\alpha_{1}}(\theta)} \left(1- \frac{8\pi c^3 \cosh^2(\alpha_1)}{15|\Omega|} \bar{I_2}P_2(\sin\theta)\right),
		\end{aligned}
	\end{equation}
	where 
	
	$$\bar{I_2} =  \frac{5i}{6} \frac{P_2'(i\sinh\alpha_{1})Q_2'(i\sinh\alpha_{2}) - Q_2'(i\sinh\alpha_{1})P_2'(i\sinh\alpha_{2})}{P_2(i\sinh \alpha_{1})Q_2'(i\sinh \alpha_{2}) - Q_2(i\sinh \alpha_{1}) P_2'(i\sinh \alpha_{2})} . $$
	In the limit $a \to 0$, an oblate spheroid reduces to a disk of radius $b$, which implies $\alpha_{1}=0$ and $c=b$. In this configuration, the first term of Eq. (\ref{eq:omega_obl}) reads 
	
	\begin{equation}
		\omega_{disk}(\boldsymbol{x} ) =  \frac{1}{4\pi c^2 \sin \theta} = \frac{1}{4\pi c \sqrt{c^2 - |\boldsymbol{x}|^2}},
	\end{equation}
	where $|\boldsymbol{x}| = \sqrt{x^2+y^2+z^2} = c \cos\theta$. This expression is identical to the classical Weber's result for the harmonic measure density on a disk of radius $c$ in the three-dimensional space (see \cite{sneddon1966mixed}, p. 64). Expectedly, this density is correctly normalized:
	
	\begin{equation}
		\int_{\Gamma} \omega_{disk}(\boldsymbol{x}) d\boldsymbol{x}
		= 2 \int_0^{2\pi} d\phi \int_0^c r \frac{1}{4\pi c
			\sqrt{c^2 - r^2}} dr = 1  ,
	\end{equation}
	where the factor 2 accounts for two facets of the disk.
	One sees that, in the disk limit, the first term of Eq. (\ref{eq:omega_obl}) is the classical result, while the second term is a correction related to the outer boundary. When the outer boundary goes to infinity, the second term vanishes. Indeed, one gets 
	\begin{equation}
		\bar{I_2}(\alpha_{1}=0) = \frac{20 \sinh\alpha_2 \cosh^2\alpha_2}{3i \sinh\alpha_2 \cosh^2\alpha_2 \ln\left(\frac{1 + i \sinh \alpha_2}{1 - i \sinh \alpha_2}\right) - 6\sinh^2\alpha_2 - 4}.
	\end{equation}
	Note that in the limit $\alpha_{2} \to \infty$ one has $\bar{I_2}(\alpha_1=0) \to -20/(3\pi)$.
	Hence, the second term in Eq. (\ref{eq:omega_obl}) vanishes due to the factor $1/|\Omega|$.
	
	As before, we focus on the probability $\omega(\boldsymbol{x})d\boldsymbol{x}=\hat{\omega}(\xi, \phi)d\xi d\phi$, where
	\begin{equation}
		\hat{\omega}(\xi, \phi) = c \cosh\alpha_{1}h_{\alpha_{1}}\omega(\boldsymbol{x}) = \frac{1+\bar{\gamma} P_2(\xi)}{4\pi}
	\end{equation}  
	with $\xi = \sin(\theta) \in (-1,1)$ and 
	\begin{equation}\label{eq:gamma_obl}
		\bar{\gamma} =  \frac{-2\cosh^2(\alpha_{1}) \bar{I_2}}{5 (\sinh\alpha_{2} \cosh^2(\alpha_{2}) - \sinh\alpha_{1}\cosh^2(\alpha_{1}))}.
	\end{equation}
	We checked numerically that $\bar{\gamma} > 0$ for all $\alpha_{1}<\alpha_2$.
	One can easily check that
	\begin{equation}
		\int_{\Gamma}\omega(\boldsymbol{x})d\boldsymbol{x} = \int_{-1}^{1}\int_{0}^{2\pi}\hat{\omega}(\xi, \phi)d\xi d\phi = 1.
	\end{equation}
	
	For large $\alpha_2$, we find
	
	\begin{align}\label{eq:gammab_asymp}
		\bar{\gamma} \approx -e^{-3 \alpha_{2}} \frac{8i \cosh ^{2}(\alpha_{1})}{3} \frac{Q_{2}^{\prime}\left(i \sinh \alpha_{1}\right)}{Q_{2}\left(i \sinh \alpha_{1}\right)} \quad (\alpha_2 \gg 1).
	\end{align}
	
	\begin{figure}[!h]
		\centering
		\includegraphics[width=0.65\linewidth]{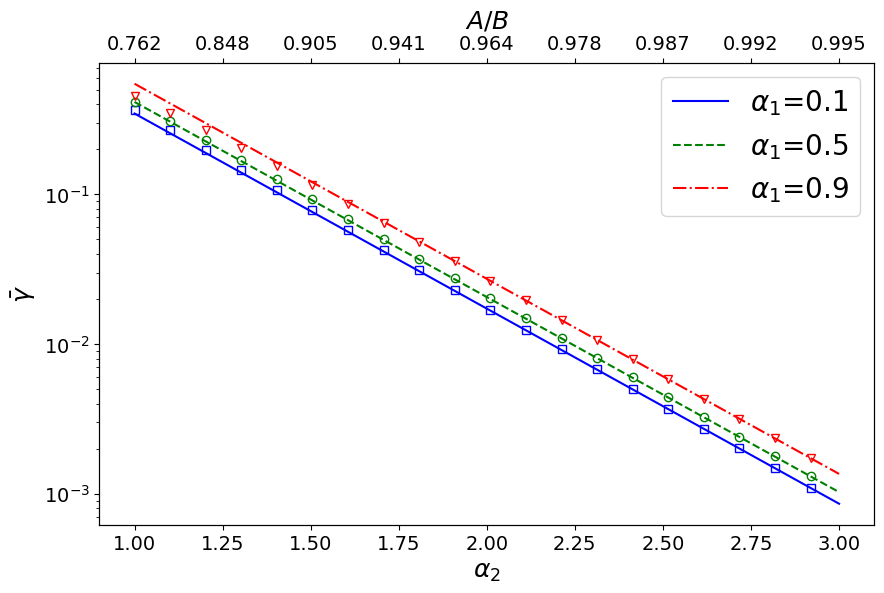}
		\caption{Semi log plot of the coefficient $\bar{\gamma}$ from Eq. (\ref{eq:gamma_obl}) (in symbols)  as a function of $\alpha_{2}$ for three values of $\alpha_{1}$, with $c=1$. Lines present the asymptotic behavior  (\ref{eq:gammab_asymp}). Note that the values of $A/B$ equivalent to $\alpha_2$ are shown on the top.}
		\label{fig:gamma_obl}
	\end{figure}
	Figure \ref{fig:gamma_obl} shows the dependence of the coefficient $\bar{\gamma}$ on the size of the domain. One sees that $\bar{\gamma}$ vanishes when the outer boundary gets larger. As a consequence, when the target is small as compared to the domain the coefficient $\bar{\gamma}$ is exponentially small that implies the uniformity of $\hat{\omega}(\xi, \phi)$.
	
	\subsection{Mean first-passage time}
	The derivation of the mean first-passage time for oblate spheroidal target is very similar.
	Using the expression (\ref{eq:G2}) of $G(\boldsymbol{x}, \boldsymbol{x}_0)$  in the oblate spheroidal coordinates we get 

	\begin{equation}
		T(\boldsymbol{x}_0) = T_0(\alpha) + P_2(\sin\theta) T_2(\alpha),
	\end{equation}
	where 
	\begin{align}\label{eq:T0_2}
		T_0(\alpha) &= \frac{c^2}{3D} \Bigl[ \frac{\sinh^2\alpha_{1} - \sinh^2\alpha}{2} + i\cosh^2\alpha_{2}\sinh\alpha_{2}( (Q_0(i\sinh\alpha_{1})-Q_0(i\sinh\alpha)) \Bigr], \\
		\nonumber T_2(\alpha) &=\frac{c^2}{9D} \biggl(1 +
		\frac{P'_2(i\sinh\alpha_2) Q_2(i\sinh\alpha)
			-Q'_2(i\sinh\alpha_2) P_2(i\sinh\alpha)}
		{P'_2(i\sinh\alpha_2) Q_2(i\sinh\alpha_1)
			-Q'_2(i\sinh\alpha_2) P_2(i\sinh\alpha_1)}
		\biggr).
	\end{align}
	
	In the limit $a \to b$ and $A\to B$, one should retrieve the mean first-passage time to a perfectly reactive spherical target of radius $\rho=a$ surrounded by a reflecting sphere of radius $R=A$ given in  Eq. (\ref{eq:tsphere}).
	Also, setting the starting position of the particle on the outer boundary, one can show that $T_0(\alpha_{2})$ exhibits the asymptotic behavior 
	
	\begin{equation}\label{eq:Tobl_as1}
		T_0(\alpha_{2}) \approx i\frac{c^2}{24D} e^{3\alpha_{2}} Q_0(i\sinh \alpha_{1}) \quad (\alpha_2\gg1),
	\end{equation} 
	that is to say, $T_0(\alpha_2)$ exponentially grows as the domain increases while $T_2(\alpha_2) \approx \cfrac{c^2}{9D}$ for $\alpha_2$ large enough. So that 
	
	\begin{equation}
		T(\boldsymbol{x}_0) \approx T_0(\alpha_2) \quad (\alpha_2\gg1).
	\end{equation}	
	Moreover, using the expression (\ref{eq:vol2}) of the volume $|\Omega|$  and the capacity $C$ of an oblate spheroid in three dimensions \cite{landau2013electrodynamics}
	\begin{equation}
		C = \cfrac{4\pi c}{\cos^{-1}\left(a/b\right)},
	\end{equation}
	we easily check the expected asymptotic relation (\ref{eq:T_as2}):
	
	\begin{equation}\label{eq:Tobl_as2}
		\overline{T} \approx \frac{|\Omega|}{D C} \approx \frac{ic^2}{24D} e^{3\alpha_{2}} Q_0(i\sinh \alpha_{1}) \quad (\alpha_2\gg1).
	\end{equation}
	Indeed, integrating $T(\boldsymbol{x}_0)$ over the volume we get 
	\begin{equation}
		\overline{T} = \frac{4\pi c^5 I}{D |\Omega|},
	\end{equation}
	with 
	\begin{align*}
		I &=\left(\sinh ^3 \alpha_2-\sinh ^3 \alpha_1\right) \frac{-12 \sinh ^2 \alpha_2+3 \sinh ^2 \alpha_1-10}{135}\\
		&+\left(\sinh \alpha_2-\sinh \alpha_1\right) \frac{\sinh ^2 \alpha_1\left(12 \sinh ^2 \alpha_2+15\right)+2}{135} \\
		&+\frac{i}{9} \cosh ^4 \alpha_2 \sinh ^2 \alpha_2\left(Q_0\left(i \sinh \alpha_1\right)-Q_0\left(i \sinh \alpha_2\right)\right) \\
		&+\frac{\cosh ^2 \alpha_1}{405 i} \frac{P_2^{\prime}\left(i \sinh \alpha_2\right) Q_2^{\prime}\left(i \sinh \alpha_1\right)-Q_2^{\prime}\left(i \sinh \alpha_2\right) P_2^{\prime}\left(i \sinh \alpha_1\right)}{P_2^{\prime}\left(i \sinh \alpha_2\right) Q_2\left(i \sinh \alpha_1\right)-Q_2^{\prime}\left(i \sinh \alpha_2\right) P_2\left(i \sinh \alpha_1\right)}.
	\end{align*}
	
\begin{figure}[!ht]
		\centering
		\includegraphics[width=0.65\linewidth]{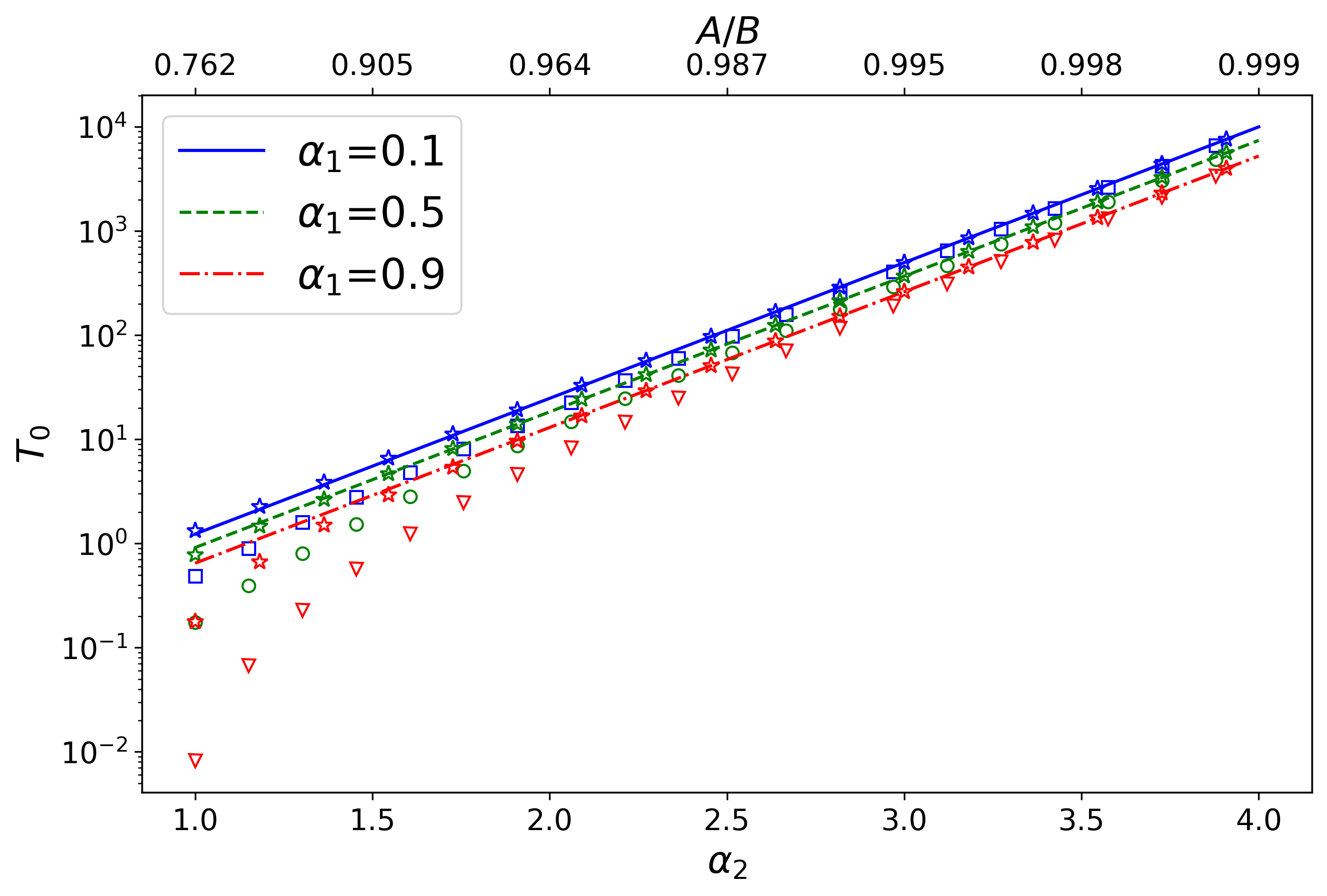}
		\caption{Semi log plot of $T_0(\alpha_{2})$ given by Eq. (\ref{eq:T0_2}) (in symbols) as a function of $\alpha_{2}$ for three values of $\alpha_{1}$, with $c=1$ and $D=1$. Lines present the asymptotic behavior  (\ref{eq:Tobl_as1}), while stars present the asymptotic relation (\ref{eq:Tobl_as2}). Note that the values $A/b$ equivalent to $\alpha_2$ are shown on top.}
		\label{fig:T0_obl}
	\end{figure}

	Figure \ref{fig:T0_obl} illustrates these asymptotic behaviors for a particle diffusing from the outer boundary. On this semi log plot, one sees the expected exponential growth of $T(\boldsymbol{x}_0)$ when the size of domain increases and the relevance of the asymptotic relations (\ref{eq:Tobl_as1}) and (\ref{eq:Tobl_as2}).
	
\begin{figure}[!ht]
		\centering
		\includegraphics[width=0.65\linewidth]{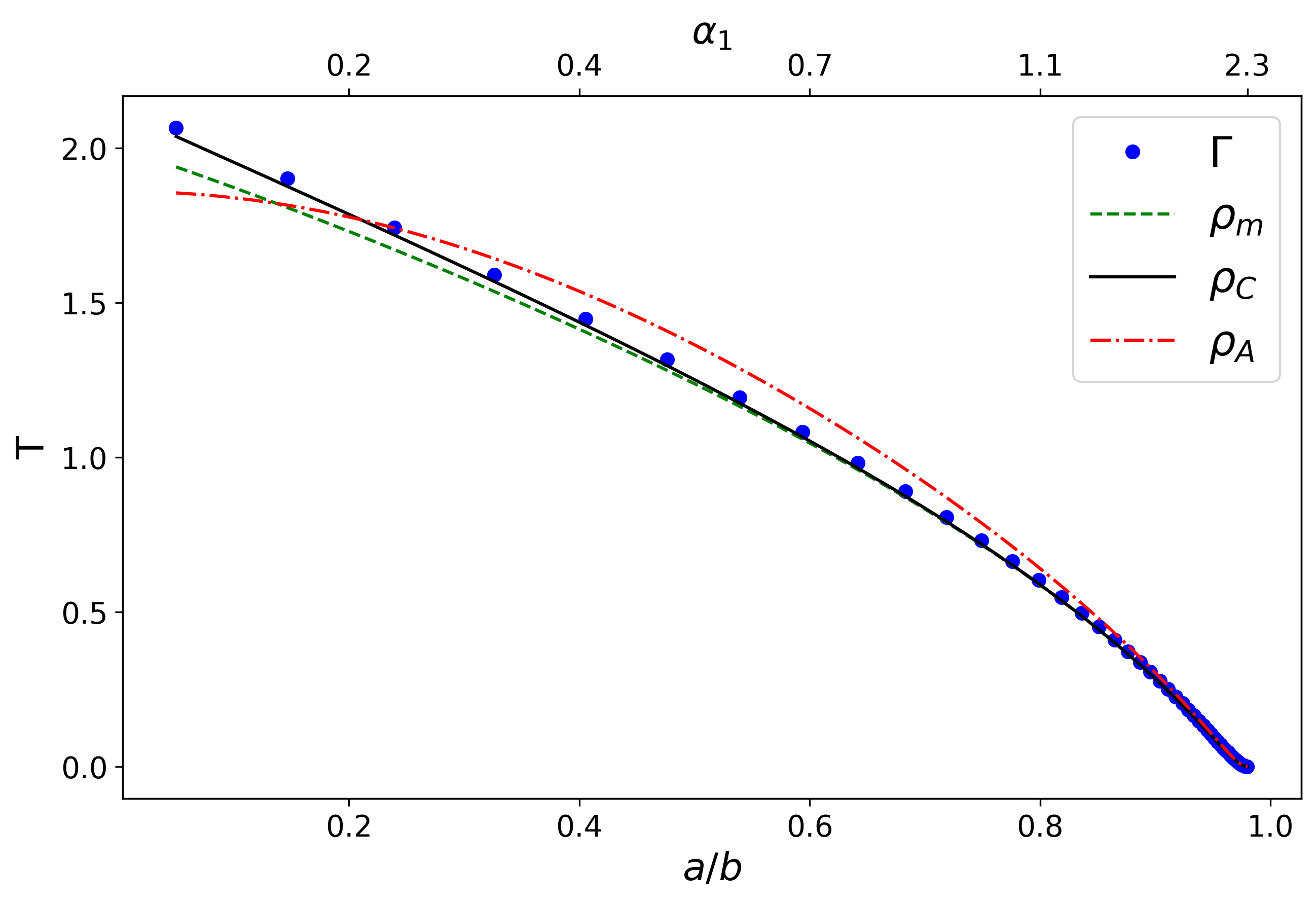}
		\caption{Mean first-passage time to an oblate spheroidal target $\Gamma$ (shown by filled circles) as a function of its aspect ratio $a/b$ for a particle diffusing from the outer boundary -- a concentric spheroid with semiaxis $A=0.99$ and $B=1.01$ which implies $\alpha_{2} = \tanh^{-1}(A/B)\approx 2.30$ and $c\approx 0.04$ according to Eq. (\ref{eq:c2}). We set $D=1$. For comparison, three curves shown by lines present the mean first-passage time to an equivalent spherical target with three choices $\rho_m$, $\rho_C$, $\rho_A$ of the effective radius.}
		\label{fig:mfpt_obl}
	\end{figure}

	In what follows, the confining spheroidal boundary $\partial \Omega_{0}$ is chosen to be close to a sphere of radius $1$, by setting $A = 0.99$ and $B = 1.01$ and thus $\alpha_{2}=\tanh^{-1}(A/B)\approx 2.30$. We compare the mean first-passage time of a particle diffusing from the outer boundary  to a spherical or to an oblate spheroidal target. As previously, we consider three criteria of ``equivalence'', by setting 
	
	\begin{align}
		\rho_m &= (a+2b)/3, \\
		\rho_C &= \frac{c}{\cos^{-1}(a/b)}, \\
		\rho_A &= \sqrt{\frac{|\Gamma|}{4\pi}}.
	\end{align}
	In the first case, the particle always reaches the sphere faster (compare blue circles and dashed green line on Figure \ref{fig:mfpt_obl}).
	In the second case,  the mean first-passage times are very close (solid black line and blue circles on Figure \ref{fig:mfpt_obl}), even when the target is small.
	In the last case, Figure \ref{fig:mfpt_obl} shows that the mean first-passage time to the anisotropic target is smaller than the mean first-passage time to the spherical target, meaning that for a given surface area the oblate spheroid presents a better ``trapping ability''. Curiously, for highly anisotropic target (e.g., a disk) the result is reversed, i.e. the mean first-passage time to the anisotropic target is greater than the mean first-passage time to the spherical target (compare blue circles and dash-dotted red line on Figure \ref{fig:mfpt_obl}).
	
	\section{Discussion and conclusion}\label{sec:discussion}
	In this paper, we investigated restricted diffusion inside a bounded domain towards a small anisotropic target. Our first result is the derivation of the exact expressions of  the volume-averaged harmonic measure density $\omega(\boldsymbol{x})$ for both prolate and oblate spheroids. The non-linear parameterization of the target surface via spheroidal coordinates strongly affects this density. In fact, the strong dependence of $\omega(\boldsymbol{x})$ on the angle $\theta$ for highly anisotropic targets emerges through the factor $1/h_{\alpha_1}(\theta)$ in Eqs. (\ref{eq:mharmofin},\ref{eq:omega_obl}) for prolate and oblate spheroids, respectively. However, this factor is compensated by $h_{\alpha_1}(\theta)$ in the surface element $d\boldsymbol{x}$. In other words, even though the volume-averaged harmonic measure density may strongly vary with $\boldsymbol{x}$, the probability $\omega(\boldsymbol{x})d\boldsymbol{x} = \hat{\omega} (\xi, \phi) d\xi d\phi$ may still be almost constant. The density $\hat{\omega}(\xi, \phi)$ depends on the anisotropy of the target through the parameter $\gamma$ (or $\bar{\gamma})$. We showed that $\gamma$ and $\bar{\gamma}$ vanish exponentially as the confining domain grows that implies the uniformity. In the case of prolate spheroids, we also showed that $\gamma$ vanishes when $\alpha_1$ goes to zero with a fixed outer boundary, so that the inaccessibility of a segment-like target to Brownian motion also restores the uniformity. Note that both options are not geometrically equivalent. Indeed, increasing  the parameter $\alpha_{2}$ results in an exponential growth of $A$ and $B$ such that the ratio $b/B$ diminishes and the fixed target is small as compared to the domain. In turn, decreasing the parameter $\alpha_{1}$ results in changes in the target shape: the semiaxis $b=c \cosh \alpha_{1}$ approaches $c$ while the other semiaxis $a=c \sinh \alpha_{1}$ vanishes. In other words, to study the effect of anisotropy of a small target, one should first fix the target and then expand the outer boundary.
	
	Our results urge to revise the uniformity hypothesis formulated in \cite{chaigneau2022first}. In fact, one derivation step consisted in replacing $\omega(\boldsymbol{x})$ by a constant $1/|\Gamma|$, which is approximately valid for moderately anisotropic targets but fails for highly anisotropic ones (see Figure \ref{fig:omega_2.15}). At the same time, our analysis of the density $\hat{\omega}(\xi, \phi)$ suggested that all points of a small spheroidal target are almost equally accessible to Brownian motion. This weaker form of the uniformity hypothesis can potentially be used to remediate the above derivation step in \cite{chaigneau2022first} and thus to extent the applicability of its results to highly anisotropic targets. Note also that irregular structure of the target surface (e.g. long channels) can fully break the equal accessibility of its points \cite{grebenkov2005makes,grebenkov2005multifractal,mandelbrot1990potential, evertsz1991behaviour}. This issue has to be investigated in the future.
	
	The second result concerned the mean first-passage time and the impact of target anisotropy that was mainly ignored in former studies. We obtained the exact formula for the mean first-passage time in a bounded domain between both prolate and oblate concentric spheroids. We illustrated the behavior of the mean first-passage time to both elongated and flattened targets.
	We showed that when the target is small as compared to the outer boundary, $T(\boldsymbol{x}_0)$  is exponentially large with respect to $\alpha_{2}$, and one retrieves the expected capacitance approximation (\ref{eq:T_as2}). 
	
	We compared the mean first-passage time of a particle diffusing from the outer boundary to a spherical or to a spheroidal target  under several criteria of ``equivalence'' between the targets. Targets with the same harmonic capacity or with the same surface area appeared to be the most interesting cases. In the former case, the mean first-passage time towards a spheroidal target is almost identical to that of the sphere, demonstrating that the capacity is indeed one of the main diffusion-sensitive characteristics of the target. Curiously, in the case of identical surface areas, the mean first-passage times to highly elongated or flattened targets (but not too flattened) are smaller than that of the spherical target. In other words, given a surface area, an anisotropic spheroid presents a better ``trapping ability'' than the sphere. 

	In summary, we provided analytical results on the effects of target anisotropy on diffusion-controlled reactions in the particular case of prolate and oblate spheroids. A more systematic analysis of these effects for small targets of arbitrary shapes presents an interesting perspective in the future.

	
	\appendix
	
	\section{Green's functions}
	Even though Green's functions are known for different sets of boundary conditions (see, e.g., \cite{chang2016green, xue2017green}), we summarize here the main formulas for our
	setting and sketch the main steps of their derivation for completeness.
	The derivations are based on an explicit representation of the Green's function in both prolate and oblate spheroidal coordinates.
	\subsection{Prolate spheroids}\label{app:pro}
	The Green's function can be decomposed in two parts and written
	\begin{equation}
		G(\boldsymbol{x}, \boldsymbol{x}_0) = \frac{1}{4\pi|\boldsymbol{x}-\boldsymbol{x}_0|} -g(\boldsymbol{x},\boldsymbol{x}_0),
	\end{equation}
	where $\frac{1}{4\pi |\boldsymbol{x}-\boldsymbol{x}_0|}$ is the fundamental solution of the Laplace equation in three dimensions and $g$ is a regular part satisfying
	
	\begin{subequations}
		\begin{align}
			\Delta g \left(\boldsymbol{x}, \boldsymbol{x}_{0}\right) =0 \quad &(\boldsymbol{x} \in \Omega), \\
			g\left(\boldsymbol{x}, \boldsymbol{x}_{0}\right) =\frac{1}{4\pi|\boldsymbol{x}-\boldsymbol{x}_0|} \quad &(\boldsymbol{x} \in \Gamma), \label{subeq:cible}\\
			\partial_{n} g\left(\boldsymbol{x}, \boldsymbol{x}_{0}\right) =\partial_n \frac{1}{4\pi|\boldsymbol{x}-\boldsymbol{x}_0|} \quad &(\boldsymbol{x} \in \partial \Omega_{0}). \label{subeq:front}
		\end{align}
	\end{subequations}
	Due to the azimuthal symmetry of the problem, the harmonic function $g(\boldsymbol{x}, \boldsymbol{x}_0)$ can be expressed as
	
	\begin{equation}
		\begin{aligned}
			g(\boldsymbol{x},\boldsymbol{x}_0) = \sum_{n=0}^{\infty} \sum_{m=-n}^{n} &P_n^m(\cos\theta_{\boldsymbol{x}}) \cos(m\phi_{\boldsymbol{x}}) P_n^m(\cos\theta_{\boldsymbol{x}_0}) \cos(m\phi_{\boldsymbol{x}_0}) \\ \times&\Bigl( C^1_{mn} P_n^m(\cosh\alpha_{\boldsymbol{x}}) + C^2_{mn} Q_n^m(\cosh\alpha_{\boldsymbol{x}}) \Bigr),
		\end{aligned}
	\end{equation}
	with coefficients $C^1_{mn}$ and $C^2_{mn}$ to be determined from boundary conditions, and $P_n^m(x)$, $Q_n^m(x)$ are the associated Legendre functions of first and second kind with $n$ and $m$ being the degree and the order. In particular, $P_n(x) = P_n^0(x)$ and $Q_n(x) = Q_n^0(x)$ are the Legendre functions of the first and second kind, respectively.

	To proceed, we use the prolate spheroidal expansion of $\frac{1}{|\boldsymbol{x}-\boldsymbol{x}_0|}$ given by \cite{morse1954methods, xue2017green}.
	One gets the coefficients $C^1_{mn}$ and $C^2_{mn}$ as

	\begin{align}
		C^1_{mn} = \frac{H_{mn}}{4\pi c\det_{mn}} Q_n^{'m}(\cosh \alpha_{2})  &\Bigl[P_n^m(\cosh \alpha_{1}) Q_n^m(\cosh \alpha_{\boldsymbol{x}_0}) - \\
		& \nonumber P_n^m(\cosh \alpha_{\boldsymbol{x}_0}) Q_n^m(\cosh\alpha_{1}) \Bigr], \\
		C^2_{mn} = \frac{H_{mn}}{4\pi c\det_{mn}}  P_n^{m}(\cosh \alpha_{1}) &\Bigl[ P_n^m(\cosh \alpha_{\boldsymbol{x}_0})Q_n^{'m}(\cosh\alpha_{2}) - \\
		&\nonumber Q_n^m(\cosh \alpha_{\boldsymbol{x}_0})P_n^{'m}(\cosh \alpha_{2}) \Bigr],
	\end{align}
	where prime denotes the derivative with respect  to the argument and
	\begin{align}
		H_{mn} &= (2n+1)(2-\delta_{m,0})(i)^m\left[ \frac{(n-m)!}{(n+m)!} \right]^2, \\
		\operatorname{det}_{mn} &= P_n^{m}(\cosh \alpha_{1})Q_n^{'m}(\cosh \alpha_{2}) - Q_n^{m}(\cosh \alpha_{1}) P_n^{'m}(\cosh \alpha_{2}).
	\end{align}
	
	In this way, for $\alpha_{\boldsymbol{x}} < \alpha_{\boldsymbol{x}_0}$, one has
	
	\begin{equation}
		\begin{aligned}
			G(\boldsymbol{x}, \boldsymbol{x}_0) = &\sum_{n=0}^{\infty} \sum_{m=-n}^{n} P_n^m(\cos \theta_{\boldsymbol{x}}) P_n^m(\cos \theta_{\boldsymbol{x}_0}) \cos(m\phi_{\boldsymbol{x}})\cos(m\phi_{\boldsymbol{x}_0})\\
			\times\Bigl[&A_{mn} P_n^m(\cosh \alpha_{\boldsymbol{x}}) P_n^m(\cosh \alpha_{\boldsymbol{x}_0}) + 
			B_{mn} Q_n^m(\cosh \alpha_{\boldsymbol{x}}) P_n^m(\cosh \alpha_{\boldsymbol{x}_0}) \\ 
			+&C_{mn} P_n^m(\cosh \alpha_{\boldsymbol{x}}) Q_n^m(\cosh \alpha_{\boldsymbol{x}_0}) + D_{mn} Q_n^m(\cosh \alpha_{\boldsymbol{x}}) Q_n^m(\cosh \alpha_{\boldsymbol{x}_0}) \Bigr],
		\end{aligned}
	\end{equation}
	
	and for $\alpha_{\boldsymbol{x}} > \alpha_{\boldsymbol{x}_0}$, one has
	
	\begin{equation}\label{eq:G}
		\begin{aligned}
			G(\boldsymbol{x}, \boldsymbol{x}_0) = &\sum_{n=0}^{\infty} \sum_{m=-n}^{n} P_n^m(\cos \theta_{\boldsymbol{x}}) P_n^m(\cos \theta_{\boldsymbol{x}_0}) \cos(m\phi_{\boldsymbol{x}})\cos(m\phi_{\boldsymbol{x}_0})\\
			\times\Bigl[&A_{mn} P_n^m(\cosh \alpha_{\boldsymbol{x}}) P_n^m(\cosh \alpha_{\boldsymbol{x}_0}) + 
			C_{mn} Q_n^m(\cosh \alpha_{\boldsymbol{x}}) P_n^m(\cosh \alpha_{\boldsymbol{x}_0}) \\ 
			+&B_{mn} P_n^m(\cosh \alpha_{\boldsymbol{x}}) Q_n^m(\cosh \alpha_{\boldsymbol{x}_0}) + D_{mn} Q_n^m(\cosh \alpha_{\boldsymbol{x}}) Q_n^m(\cosh \alpha_{\boldsymbol{x}_0}) \Bigr],
		\end{aligned}
	\end{equation}
	
	with 
	\begin{align*}
		A_{mn} &= \frac{1}{4\pi c \det_{mn}} H_{mn} Q_n^{'m}(\cosh \alpha_{2}) Q_n^m(\cosh \alpha_{1}), \\
		B_{mn} &= \frac{-1}{4\pi c \det_{mn}} H_{mn} Q_n^{'m}(\cosh \alpha_{2}) P_n^m(\cosh \alpha_{1}), \\
		C_{mn} &= \frac{1}{4 \pi c} H_{mn} - \frac{1}{4\pi c \det_{mn}} H_{mn} Q_n^{'m}(\cosh \alpha_{2}) P_n^m(\cosh \alpha_{1}), \\
		D_{mn} &= \frac{1}{4\pi c \det_{mn}} H_{mn} P_n^{'m}(\cosh \alpha_{2}) P_n^m(\cosh \alpha_{1}).
	\end{align*}
	
	From these expressions we derive the volume-averaged harmonic measure density and the mean first-passage time.
	\subsection{Oblate spheroids}\label{app:obl}
	The computation is similar for oblate spheroids. Using the oblate spheroidal expansion of $\frac{1}{|\boldsymbol{x}-\boldsymbol{x}_0|}$ given by \cite{morse1954methods}, for $\alpha_{\boldsymbol{x}} < \alpha_{\boldsymbol{x}_0}$ one has 
	\begin{equation}
		\begin{aligned}
			G(\boldsymbol{x}, \boldsymbol{x}_0) = &\sum_{n=0}^{\infty} \sum_{m=-n}^{n} P_n^m(\sin \theta_{\boldsymbol{x}}) P_n^m(\sin \theta_{\boldsymbol{x}_0}) \cos(m\phi_{\boldsymbol{x}})\cos(m\phi_{\boldsymbol{x}_0})\\
			\times \Bigl[&A_{mn} P_n^m(i\sinh \alpha_{\boldsymbol{x}}) P_n^m(i\sinh \alpha_{\boldsymbol{x}_0}) + 
			B_{mn} Q_n^m(i\sinh \alpha_{\boldsymbol{x}}) P_n^m(i\sinh \alpha_{\boldsymbol{x}_0}) \\ 
			+&C_{mn} P_n^m(i\sinh \alpha_{\boldsymbol{x}}) Q_n^m(i\sinh \alpha_{\boldsymbol{x}_0}) + D_{mn} Q_n^m(i\sinh \alpha_{\boldsymbol{x}}) Q_n^m(\sinh \alpha_{\boldsymbol{x}_0}) \Bigr],
		\end{aligned}
	\end{equation}
	
	and for $\alpha_{\boldsymbol{x}} > \alpha_{\boldsymbol{x}_0}$, one has
	
	\begin{equation}\label{eq:G2}
		\begin{aligned}
			G(\boldsymbol{x}, \boldsymbol{x}_0) = &\sum_{n=0}^{\infty} \sum_{m=-n}^{n} P_n^m(\sin \theta_{\boldsymbol{x}}) P_n^m(\sin \theta_{\boldsymbol{x}_0}) \cos(m\phi_{\boldsymbol{x}})\cos(m\phi_{\boldsymbol{x}_0})\\
			\times \Bigl[&A_{mn} P_n^m(i\sinh \alpha_{\boldsymbol{x}}) P_n^m(i\sinh \alpha_{\boldsymbol{x}_0}) + 
			C_{mn} Q_n^m(i\sinh \alpha_{\boldsymbol{x}}) P_n^m(i\sinh \alpha_{\boldsymbol{x}_0}) \\ 
			+&B_{mn} P_n^m(i\sinh \alpha_{\boldsymbol{x}}) Q_n^m(i\sinh \alpha_{\boldsymbol{x}_0}) + D_{mn} Q_n^m(i\sinh \alpha_{\boldsymbol{x}}) Q_n^m(i\sinh \alpha_{\boldsymbol{x}_0}) \Bigr],
		\end{aligned}
	\end{equation}
	with 
	
	\begin{align*}
		A_{mn} &= \frac{1}{4\pi c \det_{mn}} H_{mn} Q_n^{'m}(i\sinh \alpha_{2}) Q_n^m(i\sinh \alpha_{1}), \\
		B_{mn} &= \frac{-1}{4\pi c \det_{mn}} H_{mn} Q_n^{'m}(i\sinh \alpha_{2}) P_n^m(i\sinh \alpha_{1}), \\
		C_{mn} &= \frac{1}{4 \pi c} H_{mn} - \frac{1}{4\pi c \det_{mn}} H_{mn} Q_n^{'m}(i\sinh \alpha_{2}) P_n^m(i\sinh \alpha_{1}),\\
		D_{mn} &= \frac{1}{4\pi c \det_{mn}} H_{mn} P_n^{'m}(i\sinh \alpha_{2}) P_n^m(i\sinh \alpha_{1}).
	\end{align*}
	and 
	\begin{align}
		H_{mn} &= (2n+1)(2-\delta_{m,0})(i)^{m+1}\left[ \frac{(n-m)!}{(n+m)!} \right]^2, \\ 
		\operatorname{det}_{mn} &= P_n^{m}(i\sinh \alpha_{1})Q_n^{'m}(i\sinh \alpha_{2}) - Q_n^{m}(i\sinh \alpha_{1}) P_n^{'m}(i\sinh \alpha_{2}).
	\end{align}
	
	\newpage
	\bibliographystyle{vancouver}
	\bibliography{biblio}

\end{document}